\documentclass[fleqn,usenatbib]{mnras}

\usepackage{newtxtext,newtxmath}
\usepackage[T1]{fontenc}
\usepackage{ae,aecompl}

\usepackage[utf8]{inputenc}
\usepackage{threeparttable}
\usepackage{graphicx}
\usepackage{subcaption}
\usepackage{epstopdf}
\usepackage{threeparttable}
\usepackage[dvipsnames]{xcolor}
\usepackage{bm}
\usepackage{soul}
\usepackage{hyperref}
\usepackage{float}

\usepackage{graphicx}	% Including figure files
\usepackage{amsmath}	% Advanced maths commands
\usepackage{amssymb}	% Extra maths symbols

\newcommand{\fermi}{\emph{Fermi}}
\title[Inverse Compton Signatures of GRB Afterglows]{Inverse Compton Signatures of Gamma-Ray Burst Afterglows}

\author[H. Zhang et al.]{H. Zhang$^{1}$\thanks{E-mail: zhan2966@purdue.edu},
I.M. Christie$^{2}$\thanks{E-mail: ichristi231@gmail.com},
M. Petropoulou$^{3}$,
J.M. Rueda-Becerril$^{1}$, \&
\newauthor
 D. Giannios$^{1,4,5}$
\\
$^{1}$Department of Physics, Purdue University, 525 Northwestern Avenue, West Lafayette, IN, 47907, USA\\
$^{2}$Center for Interdisciplinary Exploration \& Research in Astrophysics (CIERA), Physics \& Astronomy, Northwestern \\ University, Evanston, IL 60208, USA \\
$^{3}$Department of Astrophysical Sciences, Princeton University, 4 Ivy Lane, Princeton, NJ 08544, USA\\
$^4$Department of Physics, University of Crete, Voutes, GR-70013, Heraklion, Greece\\
$^5$Institute of Astrophysics, Foundation for Research and Technology Hellas, Voutes, GR-70013, Heraklion, Greece
}

\date{Accepted XXX. Received YYY; in original form ZZZ}

\pubyear{2019}

\begin{document}
\label{firstpage}
\pagerange{\pageref{firstpage}--\pageref{lastpage}}
\maketitle

\begin{abstract}
The afterglow emission from gamma-ray bursts (GRBs) is believed to originate from a relativistic blast wave driven into the circumburst medium. Although the afterglow emission from radio up to X-ray frequencies is thought to originate from synchrotron radiation emitted by relativistic, non-thermal electrons accelerated by the blast wave, the origin of the emission at high energies (HE; $\gtrsim$~GeV) remains uncertain. The recent detection of sub-TeV emission from GRB~190114C by MAGIC raises further debate on what powers the very high-energy (VHE; $\gtrsim 300$GeV) emission. Here, we explore the inverse Compton scenario as a candidate for the HE and VHE emissions, considering two sources of seed photons for scattering: synchrotron photons from the blast wave (synchrotron self-Compton or SSC) and isotropic photon fields external to the blast wave (external Compton). For each case, we compute the multi-wavelength afterglow spectra and light curves. We find that SSC will dominate particle cooling and the GeV emission, unless a dense ambient infrared photon field, typical of star-forming regions, is present. Additionally, considering the extragalactic background light attenuation, we discuss the detectability of VHE afterglows by existing and future gamma-ray instruments for a wide range of model parameters. Studying GRB~190114C, we find that its afterglow emission in the \fermi-LAT band is synchrotron-dominated.The late-time \fermi-LAT measurement (i.e., $t\sim 10^4$~s), and the MAGIC observation also set an upper limit on the energy density of a putative external infrared photon field (i.e. $\lesssim 3\times 10^{-9}\,{\rm erg\,cm^{-3}}$), making the inverse Compton dominant in the sub-TeV energies.

\end{abstract}


%%%%%%%%%%%%%%%%%%%%%%%%%%%%%%%%%%%%%%%%%%%%%%%%%%

%%%%%%%%%%%%%%%%% BODY OF PAPER %%%%%%%%%%%%%%%%%%

\section{Introduction}
\label{sec:intro}
Gamma-ray bursts (GRBs) are short and intense pulses of gamma-rays that are produced by internal energy dissipation in collimated, relativistic plasma outflows launched by the collapse of massive stars \citep{Woosley1993, Paczynski1998, MacFadyen1999} or the merger of compact objects \citep{Goodman1986, Paczynski1986, Kochanek1993}. The prompt gamma-ray signal ($\sim$100 keV - 100 MeV) is followed by a broadband long-lasting emission, the so-called afterglow. This is thought to be produced by {non-thermal radiative processes of particles accelerated at} a relativistic blast wave that the outflow drives into the circumburst medium \citep{Meszaros1994,Sari1998,Dermer1998,Chiang1999,piran2004,Fan2008}.

Over the past decade the \fermi \, Large Area Telescope (LAT) has detected dozens of bursts at energies beyond 100 MeV, thus opening a new window to the electromagnetic GRB emission. The high-energy (100 MeV – 100 GeV) GRB emission usually rises quickly following the prompt keV–MeV component with a small ($\sim$ second-long) delay \citep{Omodei2009, Ghisellini2010, Ghirlanda2010} and decays with time as $\propto t^{-\chi}$ with $\chi\sim 1.2$ \citep{Zhang2011, Ackermann2013, Nava2014}. Multiwavelength observations of some GRB afterglows, for instance, GRB 130427A \citep{Kouveliotou2013}, exhibit a single spectral component from optical to multi-GeV, indicating that the origin of sub-GeV and GeV emissions can be an extension of the synchrotron component from the forward external shock \citep{Kumar2009, Ghisellini2010}. However, the emission above several GeV is incompatible with this scenario and still under debate, with possible interpretations including proton synchrotron radiation \citep{Vietri1997, Totani1998, Asano2007, Razzaque2010} or proton-induced cascades \citep{Dermer2006, Asano2007, Asano2009, Asano2010, Murase2012, Petropoulou2014}. Alternatively, gamma-ray photons can also be produced by the inverse Compton scattering of low energy seed photons from relativistic electrons accelerated at the blast wave. The seed photons can be of synchrotron origin, produced locally at the blast wave \citep[synchrotron-self Compton (SSC) models, e.g. ][]{Dermer2000, Sari2001, Zhang2001, Nakar2009} or have an external origin \citep[external Compton (EC) models, e.g. ][]{Beloborodov2005b,Fan2005,Fan2006, Wang2006,giannios2008,Beloborodov2014}.

Long-duration GRBs (LGRBs), i.e. those with durations longer than $\sim 2$~s, are believed to be associated with the death of Wolf-Rayet (WR) stars \citep{Woosley1993, MacFadyen1999, Hjorth2003}. Since its original proposition, this formation scenario has been supported by many multi-wavelength observations of LGRB host galaxies. More specifically, LGRBs are commonly found in the brighter inner regions of their hosts \citep[e.g.][]{Fruchter2006,Blanchard2016,Lyman2017}. The ultraviolet (UV) light from young stellar populations \citep{Massey1998, Crowther2007} in the star-forming regions of the host galaxy can be absorbed by interstellar dust and re-emitted in the infrared (IR) or the far-infrared (FIR). If the galaxy contains copious amounts of dust (as is the case for massive and luminous  galaxies), then nearly all of the UV starlight can be reprocessed into the IR/FIR \citep{Casey2014}. Studies of optically reddened or undetected  bursts (i.e. ``dark'' GRBs) reveal that most of the host galaxies of those dust-obscured LGRBs are  massive dusty star-forming galaxies \citep[e.g.][]{Kruhler2011,Perley2013, Perley2017,Chrimes2018}.

The presence of UV and/or IR ambient radiation fields at the explosion sites of LGRBs may have an impact on the high-energy afterglow emission. \cite{giannios2008} showed that the UV emission emitted by a massive star within the same star-forming region of the GRB progenitor, can be up-scattered by the electrons accelerated in the external shock, thereby producing a powerful gamma-ray (i.e. $1-100$~GeV) event \citep[see also][]{Lu2015}. Here, we generalize the model of \cite{giannios2008} by including the effects of EC scattering of an IR ambient photon field associated with the star-forming regions of the GRB host galaxy. By considering the IR photons, we predict more scatterings within the Thomson regime and more powerful $\sim$TeV emission, as opposed to the upscattering of UV photons. Taking into account the accompanying SSC emission, we explore the detectability of the combined Compton signals from GRB afterglows at high-energies by current and next-generation Cherenkov telescopes.

This paper is organized as follows. In Section~\ref{sec:model}, we determine the parameter regime in which the EC component dominates the high-energy afterglow emission while showing results of multi-wavelength afterglow spectra including synchrotron, SSC, and EC radiation. In Section~\ref{sec:lc}, we discuss the high-energy light curves predicted by our analytical model for both SSC-dominated and EC-dominated regimes. In Section~\ref{sec:detectability}, we discuss the effects of the extragalactic background light (EBL) attenuation on the high-energy afterglow emission and present our model predictions for the detectability of GRB afterglows by the next-generation Cherenkov Telescopes Array (CTA). Finally, in Section~\ref{sec:magic_grb_detection}, we discuss the recent MAGIC detection of GRB190114C in the context of Compton afterglow emission models. Our conclusions are provided in Section~\ref{sec:discussion}.

\section{The Multi-Wavelength Afterglow Emission}
\label{sec:model} 
In the following, we generalize the treatment of \cite{Sari2001} for the synchrotron and SSC afterglow emission by computing the Compton scattering of an ambient monochromatic photon 
field with constant energy density $U_{\rm ext}$.
In this section, we determine the parameter regime in which the EC component dominates the high-energy afterglow emission while leaving a detailed derivation of the EC afterglow spectrum in Appendix~\ref{appndx:ec}. We also show the analytical results of the multi-wavelength afterglow spectra for the synchrotron, SSC, and EC radiation.

\subsection{General Considerations}\label{sec:model-1}
We begin by considering a relativistic, adiabatic blast wave, which has relaxed into a self-similar structure, propagating through an external medium of constant number density $n$. The energy $E$ of the blast wave is constant in  time and is given by $E=16\pi\Gamma^2 R^3 n m_{\rm p} c^2/17$ \citep{Blandford1976, Sari1997}, where $R$ and $\Gamma$ are the radius and bulk Lorentz factor of the blast wave, $m_{\rm p}$ is the proton mass, and $c$ is the speed of light. Henceforth, we focus on the deceleration  phase of the blast wave, where $\Gamma \propto R^{-3/2}$. 

Photons produced when the blast wave has reached a radius $R$ are received by an observer at time
$t \approx(1+z)R/4\Gamma^2 c$ after the GRB  trigger. From the expression of the blast wave energy $E$ and the previous expression for the observer time $t$, one may solve for $R$ and $\Gamma$ as
\begin{equation}
\label{eqn:bw_radius}
R(t)=\left[\frac{17Et}{4\pi m_{\rm p} \,c \, n \, (1+z)}\right]^{1/4},
\end{equation}
and
\begin{equation}
\label{eqn:bw_Gamma}
\Gamma(t) =\left[\frac{17E(1+z)^3}{1024\pi m_{\rm p}c^5 \, n \,t^3}\right]^{1/8}.
\end{equation}

As the blast wave drives a relativistic shock into the circumburst medium, particles crossing the shock front are accelerated into a non-thermal distribution. Particle acceleration at relativistic shocks has been extensively studied by analytical and numerical means \citep[][see also \citet{sironi2015} for a recent review]{kirk2000,Achterberg2001,Spitkovsky2008}. In general, the accelerated non-thermal electron distribution can be modeled as a power-law extending between a minimum Lorentz factor $\gamma'_{\rm min}$ and a maximum one  $\gamma'_{\rm max}$ \citep[e.g.,][]{Sari1998}:
\begin{equation}
\label{eqn:injection_rate}
N_{\rm inj}(\gamma') \propto \gamma'^{-p} \;\;\rm{for}\;\; \gamma'_{\mathrm{min}}<\gamma'<\gamma'_{\mathrm{max}}.
\end{equation}
We note here that all quantities measured in the co-moving frame of the blast wave are denoted with a prime. 
Assuming that $\gamma'_{\rm max}\gg \gamma'_{\rm min}$ and $p>2$, the minimum Lorentz factor $\gamma'_{\rm min}$ of the non-thermal particle distribution can be estimated by\footnote{The case of $\gamma'_{\max}\gtrsim \gamma'_{\min}$ has been discussed in \cite{Petropoulou2011}.}
\begin{equation}
\label{eqn:gamma_min}
\gamma'_{\rm min}\approx \epsilon_{\rm e}\left(\frac{p-2}{p-1}\right)\frac{m_{\rm p}}{m_{\rm e}}\Gamma,
\end{equation}
where $\epsilon_{\mathrm{e}}$ is the fraction of the shock energy transferred into relativistic electrons \citep{Sari1998}. The maximum Lorentz factor $\gamma'_{\rm max}$ can be determined by balancing the acceleration and synchrotron loss rates \citep{deJager1996, Dermer2009}
\begin{equation}
\label{eqn:gamma_max}
\gamma'_{\max}=\left(\frac{6\pi e \epsilon_{\rm acc}}{\sigma_{\rm T}B'}\right)^{1/2},
\end{equation}
where $\epsilon_{\rm acc}\le 1$ is the ratio of acceleration rate to the maximum possible particle energy-gain rate (i.e., assuming Bohm diffusion). 
In this work, we fix $\epsilon_{\rm acc}= 0.35$.

The energy loss rates of a single electron with Lorentz factor $\gamma'\gg 1$ due to synchrotron, SSC, and EC radiation are \citep{Rybicki_lightman}
\begin{equation}
\label{eqn:syn_power_single_particle}
P'_{\rm syn}(\gamma')=\frac{4}{3}\sigma_{\rm T} c \gamma'^{ 2} U'_{\rm B},
\end{equation} 
\begin{equation}
\label{eqn:SSC_power_single_particle}
P'_{\rm SSC}(\gamma')=\frac{4}{3}\sigma_{\rm T} c \gamma'^{ 2} U'_{\rm syn},
\end{equation}
and 
\begin{equation}
\label{eqn:EC_power_single_particle}
P'_{\rm EC}(\gamma')=\frac{4}{3}\sigma_{\rm T} c \gamma'^{ 2} U'_{\rm ext},
\end{equation}
where eqns.~(\ref{eqn:SSC_power_single_particle})--(\ref{eqn:EC_power_single_particle}) are valid in the Thomson regime and $U'_{\mathrm{B}}$, $U'_{\mathrm{syn}}$, and $ U'_{\rm ext}\equiv\Gamma^2 U_{\rm ext}$ \citep{Dermer1995} are the energy densities of the magnetic field, synchrotron photons, and ambient external photons in the shocked fluid frame, respectively.  The magnetic field strength in the co-moving frame of the blast wave is written as
\begin{equation}
\label{eqn:B_field}
B'=(32\pi m_{\rm p}\epsilon_{\rm B} n)^{1/2} \, \Gamma c,
\end{equation}
where $\epsilon_{\rm B}$ is the fraction of the shocked fluid energy that is carried by the magnetic field.

The characteristic cooling timescale of an electron, with Lorentz factor $\gamma'$, due to synchrotron, SSC, and EC radiation is given by
\begin{equation}
\label{eqn:cooling_timescale}
\tau'_{\rm c} \approx \frac{\gamma' m_{\rm e}c^2}{P'_{\rm EC}+P'_{\rm syn}+P'_{\rm SSC}},
\end{equation}
while the expansion time of the blast wave is written as
\begin{equation}
\label{eqn:expansion_timescale}
\tau'_{\rm exp}\approx \frac{5R}{8\Gamma c}.
\end{equation}
By equating the two aforementioned timescales, we can estimate the characteristic cooling Lorentz factor as
\begin{equation}
\label{eqn:gamma_c}
\gamma'_{\rm c} \simeq \frac{6\Gamma m_{\rm e}c^2}{5 \sigma_{\rm T} R (U'_{\rm B}+U'_{\rm syn}+U'_{\rm ext})}, \end{equation}
which can be more conveniently expressed as
\begin{equation}
\label{eqn:gamma_c_2}
\gamma'_{\rm c}=\frac{\gamma'^{\,{\rm syn}}_{\rm c}}{1+x+y},
\end{equation}
where $x \equiv U'_{\rm syn}/U'_{\rm B}$, $y \equiv U'_{\rm ext}/U'_{\rm B}$, and the synchrotron cooling Lorentz factor is given by
\begin{equation}
\label{eqn:gamma_c_syn}
\gamma'^{\,{\rm syn}}_{\rm c}\equiv\frac{6\Gamma m_{\rm e}c^2} {5\sigma_{\rm T} R U'_{\rm B}}\approx 1800\,\epsilon_{\rm B,-2}^{-1}E_{54}^{-3/8}n_0^{-5/8}\left(\frac{t_{\rm d}}{1+z}\right)^{1/8}.
\end{equation}
Henceforth, we adopt the notation $Q_x = Q/10^x$ in cgs units and $t_{\rm d}\equiv t/{\rm 1 \, day}$. In what follows, we assume that $x$ and $y$ are dominated by their values in the Thompson regime, and discuss the effects of the Klein-Nishina (KN) suppression at the end of this section.

The ratio $U'_{\rm ext}/U'_{\rm B}$ can be written as
\begin{equation}
\label{eqn:ec_y}
y=\frac{\Gamma^2 U_{\rm ext}}{U'_{\rm B}}=\frac{U_{\rm ext}}{4 n m_{\rm p} \epsilon_{\rm B} c^2} = 0.017 \, U_{\rm ext, -6} \,\epsilon_{\rm B,-2}^{-1} \, n_0^{-1},
\end{equation}
and remains constant at all stages of the blast wave evolution.
The ratio $U'_{\rm syn}/U'_{\rm B}$, which is a measure of the SSC to synchrotron losses, can be written as \citep[see also][]{Sari2001}
\begin{equation}
\label{eqn:syn_x}
x=\frac{U'_{\mathrm{syn}}}{U'_{\mathrm{B}}}=\frac{\eta U'_{\mathrm{e}}}{U'_{\mathrm{B}}(1+x+y)}=\frac{\eta \epsilon_{\mathrm{e}}}{\epsilon_{\mathrm{B}}(1+x+y)}.
\end{equation}
Here, $U'_{\rm e}$ is the kinetic energy density of relativistic electrons and $\eta$ is the radiative efficiency, namely the fraction of the electron energy radiated away via synchrotron, SSC, and EC processes. The latter can be written as
\begin{equation}
\label{eqn:eta}
\eta = \left\{
\begin{array}{ll}
1 &\gamma'_{\rm min} > \gamma'_{\rm c}\\
\left(\dfrac{\gamma'_{\mathrm{c}}}{\gamma'_{\mathrm{min}}}\right)^{2-p} = \left[\dfrac{t}{t_0(1+x+y)^2}\right]^{\frac{2-p}{2}} &\gamma'_{\rm min} < \gamma'_{\rm c}
\end{array}
\right.,
\end{equation}
where $\gamma'_{\rm min}$ and $\gamma'_{\rm c}$ are given in eqns.~(\ref{eqn:gamma_min}) and (\ref{eqn:gamma_c_syn}), respectively, while $t_0$ is the transition time from the fast cooling (i.e. $\gamma_{\rm min}^\prime > \gamma_{\rm c}^\prime$) to the slow cooling (i.e. $\gamma_{\rm min}^\prime < \gamma_{\rm c}^\prime$) regime (considering only synchrotron losses) 
\begin{equation}
\label{eqn:t_0}
t_0\approx1.2 \left(\frac{p-2}{p-1}\right)^2\epsilon_{\rm e,-1}^{2}\epsilon_{\rm B,-2}^{2}E_{54}\,n_0 \, (1+z)~{\rm d}.
\end{equation}
Substitution of eqn.~(\ref{eqn:syn_x}) to eqn.~(\ref{eqn:eta}) yields
\begin{equation}
\label{eqn:xy}
\begin{array}{ll}
x(1+x+y)=\dfrac{\epsilon_{\mathrm{e}}}{\epsilon_{\mathrm{B}}} &\gamma'_{\rm min} > \gamma'_{\rm c}\\
x(1+x+y)^{3-p}= \dfrac{\epsilon_{\mathrm{e}}}{\epsilon_{\mathrm{B}}} \left(\dfrac{t}{t_0}\right)^{(2-p)/2} &\gamma'_{\rm min} < \gamma'_{\rm c}.
\end{array}
\end{equation}
Depending on the ordering of $x$ and $y$, one can define two regimes of particle cooling and Compton emission:
\begin{itemize}
\item SSC-dominated, for $x\gg y>1$ \citep[see][for numerical results]{Petropoulou2009}. Here, $x$ is given by
\begin{equation}
x\simeq\left\{
\begin{array}{ll}
\sqrt{\dfrac{\epsilon_{\mathrm{e}}}{\epsilon_{\mathrm{B}}}} &\gamma'_{\rm min} > \gamma'_{\rm c}\\
\left(\dfrac{\epsilon_{\mathrm{e}}}{\epsilon_{\mathrm{B}}}\right)^{\frac{1}{4-p}} \left(\dfrac{t}{t_0}\right)^{\frac{2-p}{2(4-p)}} &\gamma'_{\rm min} < \gamma'_{\rm c}
\end{array}
\right .
\end{equation}
\item EC-dominated, for $y\gg x>1$. Here, $x$ is given by
\begin{equation}
x\simeq\left
\{\begin{array}{ll}
\dfrac{1}{y}\dfrac{\epsilon_{\mathrm{e}}}{\epsilon_{\mathrm{B}}} &\gamma'_{\rm min} > \gamma'_{\rm c}\\
\dfrac{1}{y^{3-p}}\dfrac{\epsilon_{\mathrm{e}}}{\epsilon_{\mathrm{B}}} \left(\dfrac{t}{t_0}\right)^{\frac{2-p}{2}} &\gamma'_{\rm min} < \gamma'_{\rm c}
\end{array}
\right..
\end{equation}
\end{itemize}
In  both the SSC-dominated and EC-dominated cooling regimes, we find that $x$ is independent of time in the fast cooling regime, but it decreases gradually once the system enters the slow cooling regime (this is valid for $p \sim 2.1-2.5$). 

\begin{figure}
\includegraphics[width=\columnwidth]{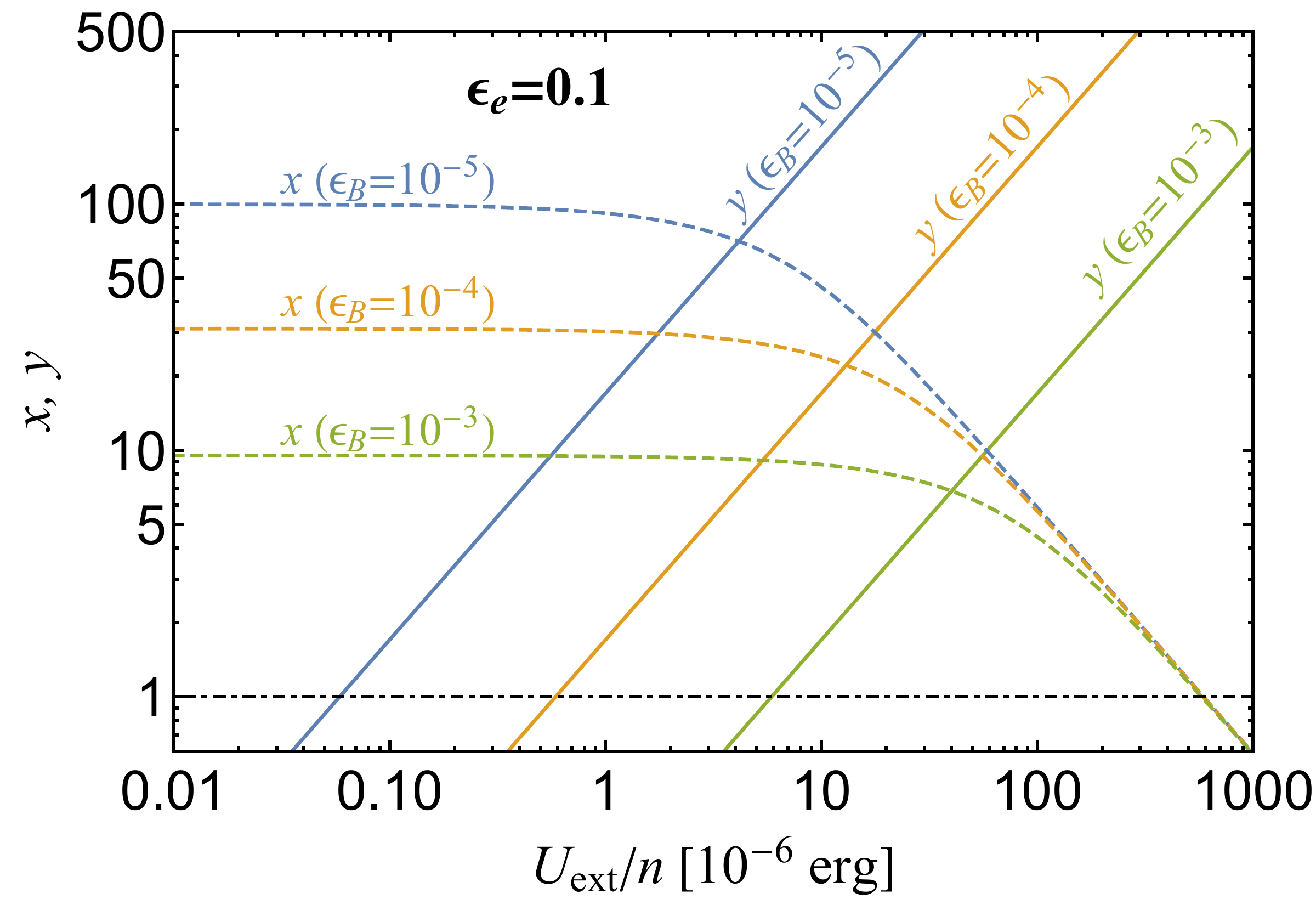}
\caption{Plot of $x\equiv U'_{\rm syn}/U'_{\rm B}$ and $y\equiv U'_{\rm ext}/U'_{\rm B}$ as a function of $U_{\rm ext}/n$ for $\epsilon_{\rm e}=0.1$ and different values of $\epsilon_{\rm B}$ as marked on the plot. The results are applicable to the fast cooling regime (i.e. $\gamma_{\rm c}^\prime < \gamma_{\rm min}^\prime$, see eqn.~\ref{eqn:xy}). For large ratios of $U_{\rm ext}/n$, the external Compton component dominates (i.e. $y\gg x$). Small values of $U_{\rm ext}/n$  (e.g. $U_{\rm ext}/n\lesssim10^{-5}$~erg for $\epsilon_{\rm B}=10^{-4}$) corresponding to an SSC-dominated cooling scenario (i.e. $x\gg y$). Below the horizontal dash-dotted line, synchrotron radiation dominates the particle cooling (i.e. $x,y<1$). A coloured version of this plot is available online.}
\label{fig:xy}
\end{figure}

Fig.~\ref{fig:xy} shows the dependence of $x$ and $y$ on $U_{\rm ext}/n$ for different values of $\epsilon_{\rm B}$  according to eqns.~(\ref{eqn:ec_y}) and~(\ref{eqn:xy}). For illustration purposes, we consider only the fast cooling regime while noting that the temporal dependence of $x$ in the slow cooling regime is weak for $p\sim2$. SSC dominates electron energy losses (i.e. $x>y$) in the fast cooling regime, if the following condition is satisfied
\begin{equation}
\label{eqn:criticalcondition}
\epsilon_{\rm e,-1}^{1/2}\,\epsilon_{\mathrm{B},-2}^{1/2}\,n_0\,U_{\mathrm{ext},-6}^{-1}\gtrsim 5.4\times10^{-3}.
\end{equation}

So far, we have assumed that inverse Compton scattering (both SSC and EC) takes place in the Thomson regime. However, KN suppression may change significantly the effective values of $x$ and $y$. The effects of KN scatterings on $x$, representing the Compton parameter of SSC, have been fully investigated by \citet{Nakar2009}. As for $y$, the Compton parameter of EC, we derive its expression including KN effects, in  Appendix~\ref{appndx:yKN}.

The KN suppression of the cross section does not only affect the values of $x,y$, but also makes them dependent on the electron Lorentz factor $\gamma'$. This may lead to strong spectral features on both synchrotron and inverse Compton components \citep[][see also next section]{Moderski2005}. When KN effects are taken into account, eqn.~\eqref{eqn:gamma_c_2} is rewritten as:
\begin{equation}\label{eqn:gamma_c_3}
    \gamma'_{\rm c}(1+x(\gamma'_{\rm c})+y(\gamma'_{\rm c}))=\gamma_{\rm c}'^{\,{\rm syn}},
\end{equation}
where the values of $x(\gamma'_{\rm c})$ and $y(\gamma'_{\rm c})$ are given by relevant equations in Section 3 and Eqn. 46 in \citet{Nakar2009} and Eqn.~\eqref{eqn:ykn} in Appendix~\ref{appndx:yKN}.

Eqn.~\eqref{eqn:gamma_c_3} can be simplified to eqn.~\eqref{eqn:gamma_c_2},
when $\Gamma\gamma'_{\rm c}\epsilon_0\ll m_{\rm e}c^2$ and $\gamma'_{\rm c}\nu_{\rm syn}(\gamma'_{\rm c})\ll\Gamma m_{\rm e}c^2$. Under such conditions, $\gamma'_{\rm c}$ is not affected by KN suppression. Otherwise, it should be solved numerically from eqn.~(\ref{eqn:gamma_c_3}).

\subsection{Multi-Wavelength Afterglow Spectra}\label{sec:model-2}

The synchrotron and SSC spectra have been extensively discussed in the literature \citep[see, e.g. ][for details]{Sari1998, Sari2001}. Analytical expressions for the EC emission of the afterglow are provided in Appendix~\ref{appndx:ec}. For the following illustrative examples, we consider an external monochromatic photon field of energy $\epsilon_{\rm 0}\sim 8 \times 10^{-3}$~eV,  as expected from dust heated to $T \simeq 90$~K \citep{wilson2014,scoville2015,Yoast-Hull2015,Perley2017,Yoast-Hull2017}. All other parameters are listed in Table~\ref{tab:parameters}.

\begin{figure}
    \centering
    \begin{subfigure}{\columnwidth}
       \includegraphics[width=1\linewidth]{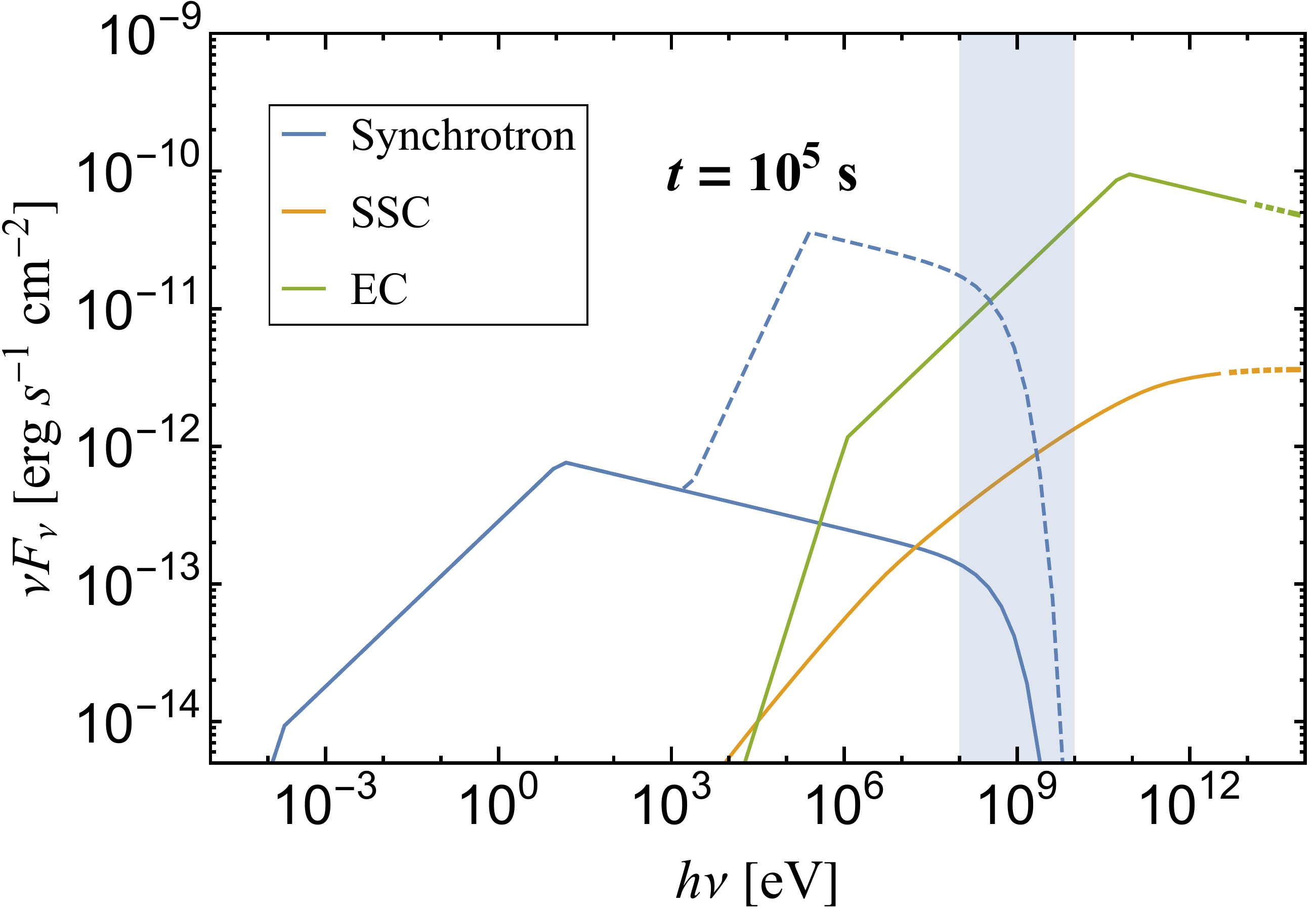}
       \caption{EC-dominant}
       \label{fig:spectra_ec} 
    \end{subfigure}

    \begin{subfigure}{\columnwidth}
       \includegraphics[width=1\linewidth]{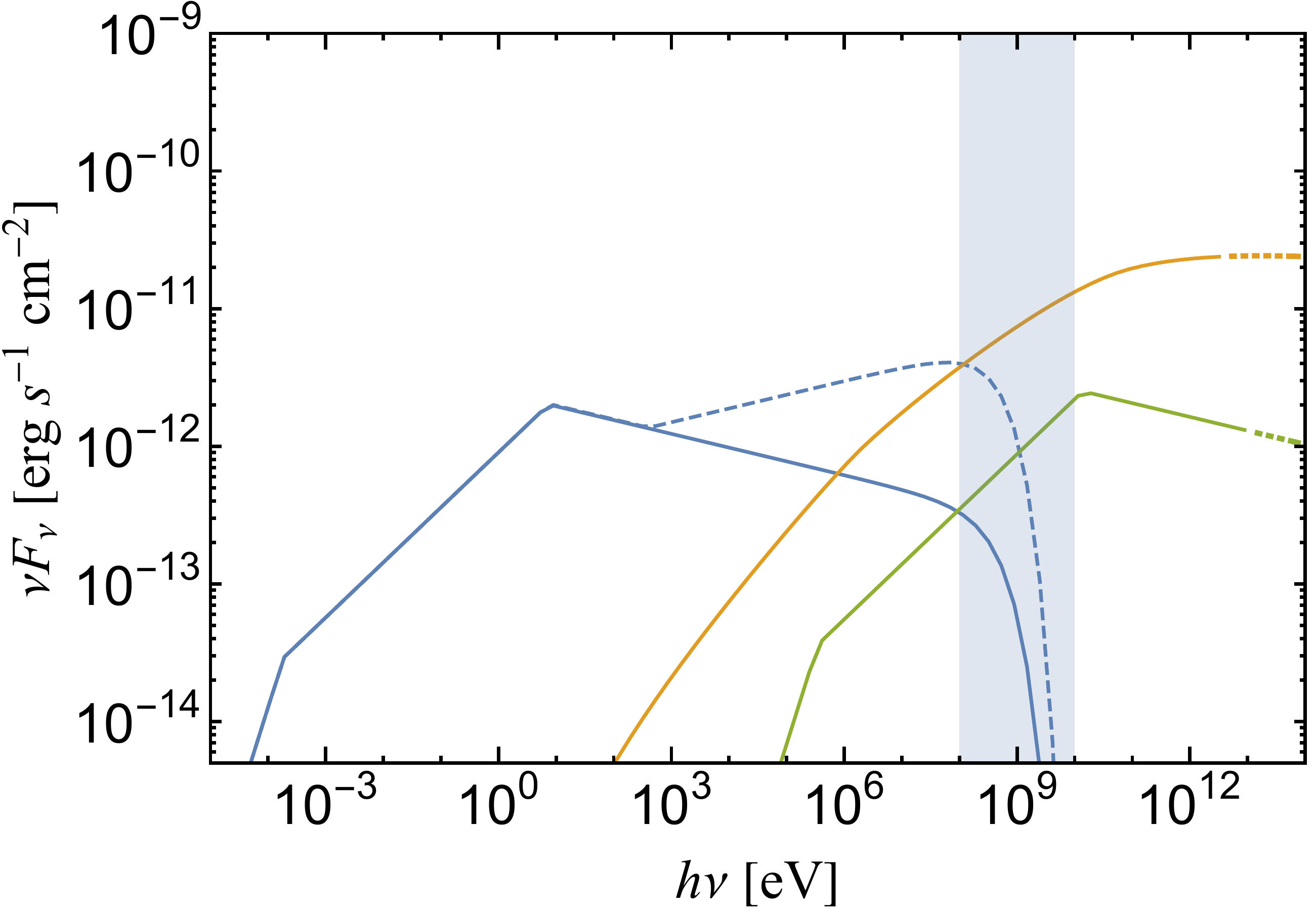}
       \caption{SSC-dominant}
       \label{fig:spectra_ssc}
    \end{subfigure}
    \caption{
    Multi-wavelength spectra of the synchrotron, SSC, and EC emission (see legend in top panel), computed at $t = 10^5$~s using the parameters listed in Table~\ref{tab:parameters}. Panels (a) and (b) show results in the EC-dominated and SSC-dominated regimes, respectively. The dashed blue lines demonstrate the Klein-Nishina effect on synchrotron spectra. The part of the Compton spectrum which should be affected by KN suppresion is shown with a dotted line. The grey-colored region indicates the 0.1-10 GeV \textit{Fermi} energy band. In both examples, the low-energy part of the spectrum is dominated by synchrotron emission, peaking at $\sim 10$~eV. The transition from the SSC-dominated (panel b) to the EC-dominated (panel a) case is achieved by increasing $U_{\rm ext} / n$ by two orders of magnitude (see eqn.~\ref{eqn:criticalcondition}).  The attenuation of high-energy gamma-rays, due to $\gamma \gamma$-absorption by the EBL, is not included here (see Sec.~\ref{sec:detectability} for more details). A coloured version of this plot is available online.}
    \label{fig:spectra}
\end{figure}

Multi-wavelength spectra, including synchrotron, SSC, and EC emission, are shown in Fig.~\ref{fig:spectra} for an observer time $t = 10^5$~s. Panels (a) and (b) show examples of the EC-dominated and SSC-dominated cases, respectively. The transition from the latter to the former regime is achieved by increasing the ratio $U_{\rm ext}/n$ (see also eqn.~\ref{eqn:criticalcondition}) by two orders of magnitude. This effectively results in an increase of the EC flux by a factor of $\sim 20$ (see eqn.~\ref{eqn:fmax_ec}). For a summary of the parameters values used in Fig.~\ref{fig:spectra}, see Table~\ref{tab:parameters}. 
 
We define two characteristic observed frequencies of the synchrotron spectra, namely 
\begin{equation}\label{eq:synmin}
\nu_{\rm min} \equiv \Gamma\gamma'^2_{\rm min} \frac{e B'}{2\pi m_{\rm e}c}
\end{equation}
and
\begin{equation}\label{eqn:syncoolingbreak}
\nu_{\rm c} \equiv \Gamma\gamma'^2_{\rm c} \frac{e B'}{2\pi m_{\rm e}c}.
\end{equation}
For the EC-dominated case (Fig.~\ref{fig:spectra_ec}), we find $h\nu_{\rm min}\simeq1.8\times 10^{-4}$~eV and $h\nu_{\rm c}\simeq 12$~eV, while for the SSC-dominated case (Fig.~\ref{fig:spectra_ssc}), the peak of the synchrotron spectrum occurs at $h\nu_{\rm c}\simeq 7$~eV; for the adopted parameter values (see Table~\ref{tab:parameters}), the minimum synchrotron frequency is the same as in the EC-dominated case.  

\begin{table}
    \centering
    \caption{Model parameters used for the indicative examples of multi-wavelength afterglow emission shown in Figs.~\ref{fig:spectra} and \ref{fig:lightcurves_all}. The minimum and maximum Lorentz factors of the electron distribution can be obtained from eqns.~(\ref{eqn:gamma_min}) and (\ref{eqn:gamma_max}), respectively.}
    \label{tab:parameters}
    \begin{tabular}{ccc}
        \cline{1-3}
        Parameters and units                 & EC-dominated      & SSC-dominated         \\
        \cline{1-3}
        $n$ [cm$^{-3}$]                      & $0.1$               & $1$                 \\
        $U_{\rm ext}$ [${\rm erg\,cm^{-3}}$] & $7.5\times 10^{-7}$ & $7.5\times 10^{-8}$ \\
        $\epsilon_{0}$ [eV]                  & $0.02$              & $0.02$              \\
        $E$ [erg]                            & $10^{54}$           & $10^{54}$           \\ 
        $\epsilon_{\rm e}$                   & $0.1$               & $0.1$               \\
        $\epsilon_{\rm B}$                   & $10^{-5}$           & $10^{-5}$           \\
        $\epsilon_{\rm acc}$               & $0.35$               & $0.35$              \\    
        $p$                                  & $2.2$               & $2.2$               \\
        % $z$                                  & $0.5$               & $0.5$               \\
        $d_{\rm L}$ [cm]                     & $9\times 10^{27}$   & $9\times 10^{27}$   \\
        \cline{1-3}
    \end{tabular}
\end{table}

In both panels, we show computed spectra from our analytical model\footnote{The SSC spectrum appears smooth due to numerical integration of the Compton emissivity over the seed synchrotron photon spectrum.}. The temporal evolution of the spectra, for both cases, can be found online\footnote{\url{https://drive.google.com/open?id=1-JAk6S3FOVU7Zz9irdtEmIB5P1OOCa2b}}, where we find that all fluxes decrease with time, yet SSC drops slightly faster than synchrotron and EC. The KN suppression of the Compton scattering cross section is not included in our analytical treatment, but it is expected to affect the part of the Inverse Compton spectrum highlighted with dotted lines. More specifically, in the SSC spectrum, the KN effects become important above $2\nu_{\rm c}\gamma_{\rm KN}^{'2}$ \citep[see Sec.~4 and eqn.~50 in][]{Nakar2009}, where $\gamma'_{\rm KN}$ is the Lorentz factor of electrons which can upscatter photons with $h\nu_{\rm c}$ in the KN regime, i.e. $h\nu_{\rm c}\gamma'_{\rm KN}/\Gamma\sim m_{\rm e}c^2$ \citep[see Eqn.~6 in][]{Nakar2009}. For the parameter values used in Fig.~\ref{fig:spectra}, we estimate that the KN effects on the SSC spectra at that time become apparent above $\sim 5$~TeV. For the EC spectrum, the KN cutoff becomes relevant at even higher energies (here, $\sim 10$~TeV) -- see also  eqn.~(\ref{eqn:EC_KN_cutoff}). 

Besides the spectral steepening of the inverse Compton component, as discussed above, the KN suppression can have a substantial impact on the synchrotron spectrum,  because it also affects the electron cooling, as discussed in the previous section. Qualitatively speaking, electrons that are up-scattering photons predominantly in the KN regime are cooling less efficiently due to inverse Compton scattering, and can radiate away their energy via synchrotron instead \citep{Moderski2005}. To illustrate this in a quantitative way, we show in Fig.~\ref{fig:spectra} the synchrotron spectra after accounting for the KN effects in electron cooling (dashed blue lines). The enhancement of synchrotron flux at energies well above that of the cooling break (here, at $1$~keV) can potentially change the model prediction in the \fermi-LAT energy range by more than one order of magnitude.

For the EC-dominated case, the ``jump'' in the synchrotron spectrum happens at a frequency that corresponds to radiating electrons with $\gamma'^*\simeq m_{\rm e}c^2/\Gamma\epsilon_0$. Although these electrons up-scatter photons of energy $\epsilon_0=h\nu_0$ in the KN regime, they can still cool down by Thomson-scattering off lower energy photons, i.e., from the Rayleigh-Jeans part of the external photon spectrum. As the relevant photon energy density decreases for electrons with $\gamma^\prime >\gamma'^*$, so does the cooling efficiency via Compton scattering. This explains the sharp enhancement of the synchrotron spectrum. It is also noteworthy that the frequency where the flux enhancement happens does not depend on time: $h\nu=\gamma'^{*2}\Gamma h e B'/2\pi m_{\rm e} c\approx30 \sqrt{n_0\epsilon_{\rm B,-4}}\,{\rm keV}$. Thus, a hard synchrotron spectrum at above $30\sqrt{n_0\epsilon_{\rm B,-4}}$~keV might be a signature of external Compton scattering. In the SSC-dominated case, the photon field is synchrotron radiation, which is much softer than Rayleigh-Jeans spectrum. We therefore expect a much softer transition, as shown in  Fig.~\ref{fig:spectra_ssc}.

\section{Gamma-ray Light Curves}
\label{sec:lc}

The high energy emission (100~MeV - 100~GeV) of GRB afterglows has been found to peak after the prompt keV-MeV component within seconds, and then decays as $\propto t^{-\chi}$, with $\chi\sim 1.2$ \citep{Ghisellini2010, Ghirlanda2010, Ackermann2013}. 
Here, we explore the temporal trends predicted in our analytical model for both SSC and EC dominated regimes.

\begin{figure}
	\includegraphics[width=\columnwidth]{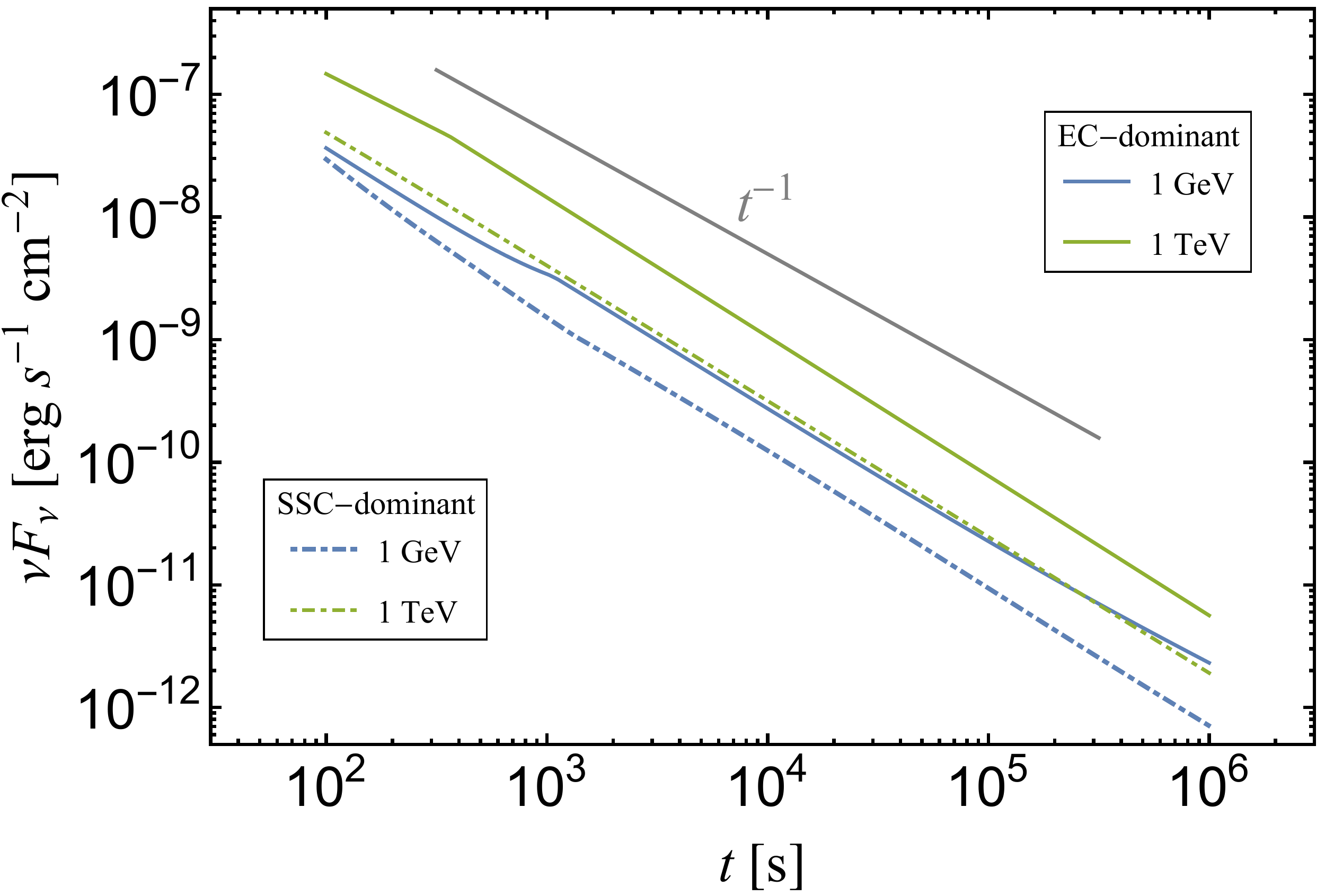}
	\caption{GRB afterglow light curves (total emission, including synchrotron, SSC, and EC), at $1$~GeV (blue) and $1$~TeV (green), as produced from our analytical calculations for the parameters listed in Table~\ref{tab:parameters}. The light curves from both the SSC and EC dominated regimes, represented by the two different line types (see legends), follow a temporal decay of $\sim t^{-1}$ (see black line for visual reference), resembling those found in GRB afterglow light curves by \fermi-LAT \citep{Ackermann2013}. A coloured version of this plot is available online. }
    \label{fig:lightcurves_all}
\end{figure}

As an indicative example, we show in Fig.~\ref{fig:lightcurves_all} the $1$~GeV and $1$~TeV light curves, which are computed using all possible contributions  from synchrotron, SSC, and EC, for the same parameters used in Fig.~\ref{fig:spectra} (see also Table~\ref{tab:parameters}). The flux at a fixed frequency decays as a single power law in time (i.e. $F_{\nu}\propto t^{-\chi}$), as long as it is produced by a single emission mechanism (either EC or SSC). The broken power-law light curve obtained at 1~GeV for both SSC- and EC-dominated cases is the result of the transition from a synchrotron-dominated to an EC-dominated emission at $t\sim 10^3$~s. 

For the adopted value  of $p = 2.2$ for the electron power-law index, we find decay slopes  $\chi\sim 0.9-1.15$, which are similar to those observed in \fermi-LAT GRB light curves. Interestingly, for $p\gtrsim 2$, the predicted values of $\chi$ do not seem to depend either on the cooling regime or the origin of seed photons for Compton scattering. Our results suggest that the gamma-ray light curve alone may not be sufficient to distinguish between EC and SSC processes, and multi-wavelength spectral and temporal information is thereby required to identify the dominant mechanism.

To further expand upon this, we present parametric scalings of the observed flux on the model parameters. In the EC dominated regime, the inverse Compton flux scales as (see also eqns.~\ref{eqn:slowspectrum1b} and \ref{eqn:fastspectrum1c} in Appendix~\ref{appndx:ec})
\begin{eqnarray}
\label{eqn:EC_flux_scaling}
F_{\nu}\propto\left\{
 \begin{array}{ll}
U_{\mathrm{ext}} E^{\frac{p+3}{4}} n^{-\frac{p-1}{4}} \epsilon_{\rm e}^{p-1} \nu^{-\frac{p-1}{2}} t^{-\frac{3(p-1)}{4}} &  \nu^{\rm EC}_{\min}<\nu<\nu^{\rm EC}_{\rm c} \\
E^{\frac{p+2}{4}} n^{-\frac{p-2}{4}} \epsilon_{\rm e}^{p-1} \nu^{-\frac{p}{2}} t^{-\frac{3p-2}{4}} & \nu>\nu^{\rm EC}_{\rm c}>\nu^{\rm EC}_{\min} \\
\end{array}
\right.,
\end{eqnarray}
where the EC cooling frequency is defined as $\nu_{\rm c}^{\rm EC}\equiv\frac{4}{3}\Gamma^2\gamma_{\rm c}'^2\nu_0$ and $\nu_{\rm min}^{\rm EC}\equiv\frac{4}{3}\Gamma^2\gamma_{\rm min}'^2\nu_0$, with $\nu_0$ being the frequency of mono-chromatic external photons. Similarly, the scaling for the SSC-dominated case reads
\begin{equation}
\label{eqn:SSC_flux_scaling}
\resizebox{\linewidth}{!}{$F_{\nu}\propto\left\{
\begin{array}{ll}
E^{\frac{3p+7}{8}}n^{\frac{-p+11}{8}}\epsilon_{\rm e}^{2(p-1)}\epsilon_{\rm B}^{\frac{p+1}{4}}\nu^{-\frac{p-1}{2}}t^{\frac{-9p+11}{8}} & \nu^{\rm SSC}_{\min}<\nu<\nu^{\rm SSC}_{\rm c} \\ \\
E^{\frac{2p -3p^2 +24}{32-8p}}n^{\frac{p^2 - 14p + 24}{32 - 8p}}\epsilon_{\rm e}^{\frac{-2p^2+8p-6}{4-p}}\epsilon_{\rm B}^{\frac{-p^2+3p+2}{16-4p}}\nu^{-\frac{p}{2}}t^{\frac{9p^2 - 38p + 24}{32 - 8p}} & \nu>\nu^{\rm SSC}_{\rm c}>\nu^{\rm SSC}_{\min} \\
\end{array}
\right. ,$}
\end{equation}
where $\nu_{\min}^{\rm SSC}\equiv 2\gamma_{\rm min}'^2\nu_{\rm min}$, $\nu_{\rm c}^{\rm SSC}\equiv2\gamma_{\rm c}'^2\nu_{\rm c}$ (see eqn.~\ref{eqn:gamma_c_2}), $\nu_{\rm min}$ is the minimum synchrotron frequency as defined in \citet{Sari1998}, and $\nu_{\rm c}$ is the cooling synchrotron frequency given by eqn.~(\ref{eqn:syncoolingbreak}). 

\citet{Nava2014} considered the GeV light curves of ten GRBs detected by \fermi-LAT and found that all decay as a power-law with a similar slope, i.e. $F_{\nu}\propto t^{-1.2}$. After re-normalizing the integrated LAT luminosity to the burst's total isotropic prompt emission energy, \citet{Nava2014} showed that the light curves of all GRBs in their sample overlapped. They argued that this result supports the interpretation of the LAT emission as synchrotron radiation from external shocks. 

Here, we examine the dependence of inverse Compton emission on the total energy of the burst. In our model, the dependence of SSC and EC fluxes on $E$ is given by eqns.~(\ref{eqn:EC_flux_scaling}) and (\ref{eqn:SSC_flux_scaling}). For instance, when $p=2.2$, eqns.~(\ref{eqn:EC_flux_scaling}) and (\ref{eqn:SSC_flux_scaling}) show that the flux is proportional to $E^{1.3}\;(\nu<\nu_{\rm c}^{\rm EC})$  and $E^{1.15}\;(\nu>\nu_{\rm c}^{\rm EC})$ for EC emissions and $E^{1.7}\;(\nu<\nu_{\rm c}^{\rm SSC})$ and $E^{0.96}\;(\nu>\nu_{\rm c}^{\rm SSC})$ for SSC. We therefore find an almost linear dependence of the flux on $E$ if the LAT emission is attributed to EC scattering (independent of the cooling break) or to SSC for $\nu > \nu_{\rm c}^{\rm SSC}$.

\section{Detectability of afterglow emission at very high energies}
\label{sec:detectability}

A very high-energy (VHE; $\epsilon_{\gamma}\gtrsim 100$~GeV) detection of a GRB afterglow can be used to probe the extragalactic background light (EBL). From the far-infrared to the visible and UV wavelengths, the EBL is thought to be dominated by starlight, either through direct emission or through absorption and re-radiation by dust. These low-energy ambient photons interact with VHE photons from extragalactic sources to produce electron-positron pairs \citep{Gould1967, Puget1976}. If the redshift and the intrinsic VHE spectrum of the source are both known, then the observed spectrum can be used to constrain different EBL models. 

\begin{figure}
    \centering
    \begin{subfigure}{\columnwidth}
       \includegraphics[width=1\linewidth]{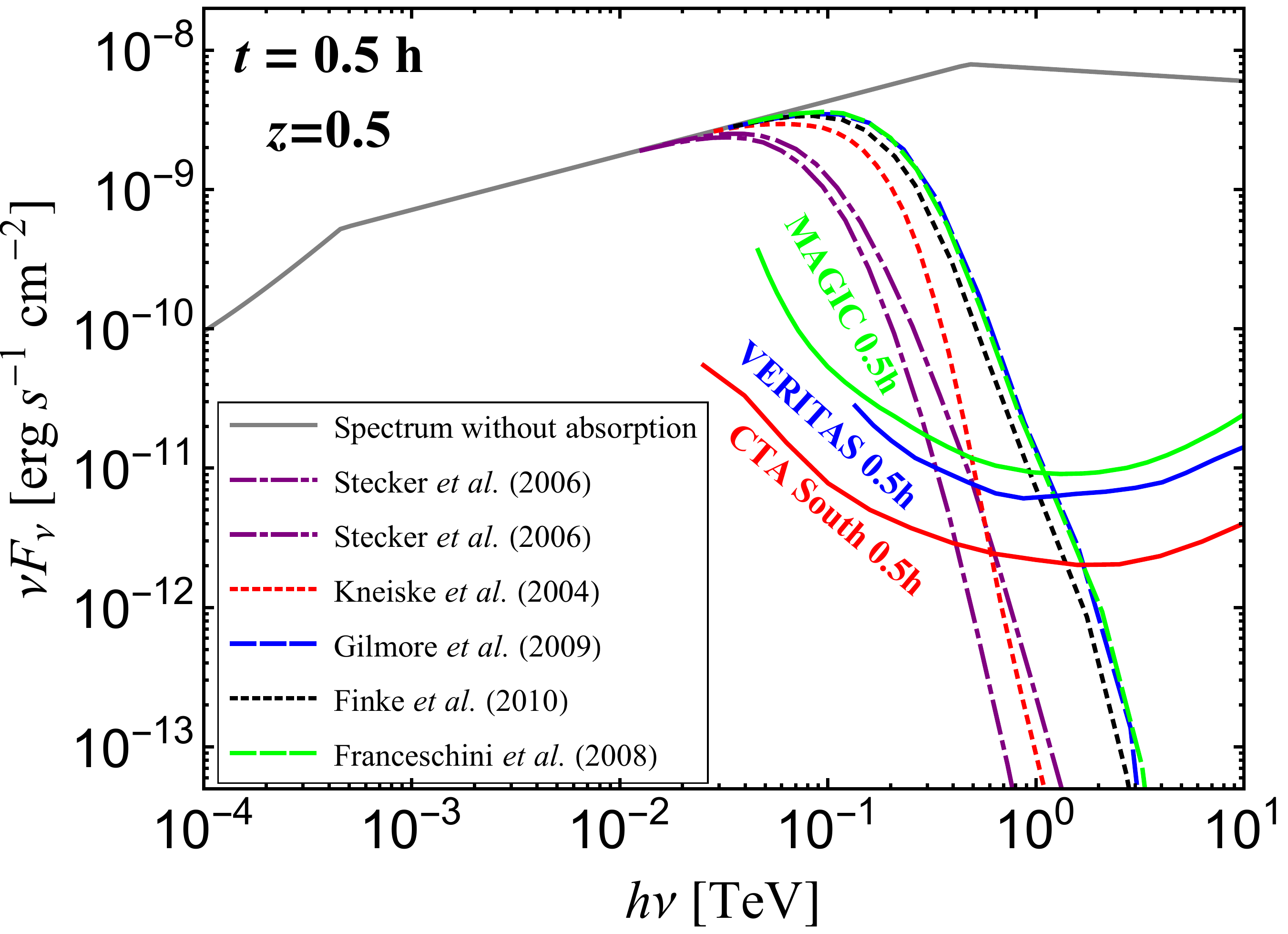}
       \label{fig:spectra_ebl0.5} 
    \end{subfigure}
    \begin{subfigure}{\columnwidth}
       \includegraphics[width=1\linewidth]{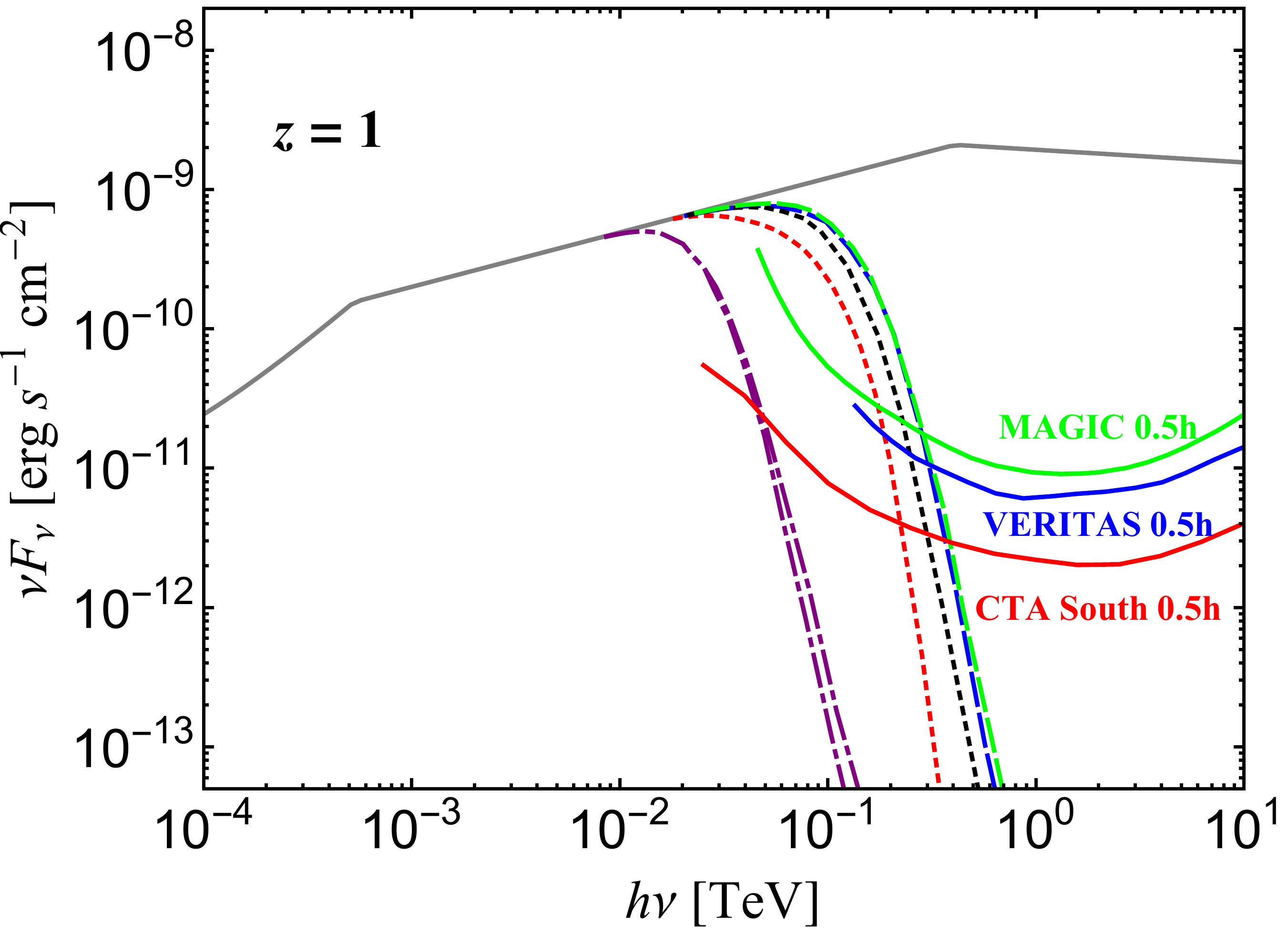}
       \label{fig:spectra_ebl1}
    \end{subfigure}
    \caption{High-energy spectrum of a GRB afterglow at $t=0.5$~hr from our EC-dominated model (see Fig.~\ref{fig:spectra_ec}) for two indicative redshifts:  $z=0.5$ (top panel) and $z=1$ (bottom panel). The gray solid line shows the spectrum without EBL absorption. The attenuated gamma-ray spectra for different EBL models \citep{Stecker2006, Kneiske2004, Gilmore2009, Finke2010, Franceschini2008} are overplotted with different coloured lines (see inset legend). For both redshifts, the EBL absorption becomes important at energies $>100$~GeV. The 0.5-hour differential flux sensitivity curves of CTA South, MAGIC, and VERITAS (overplotted with solid red, green, and blue lines, respectively) show that this event is well within the detecting capabilities of these instruments. If the intrinsic spectrum is known, its shape close to its peak energy can place strong constraints on the EBL models. A coloured version of this plot is available online.}
    \label{fig:spectrum_ebl}
\end{figure}

Fig.~\ref{fig:spectrum_ebl} shows the instantaneous VHE afterglow spectrum computed at $t=0.5$~hr with (coloured lines) and without (solid grey line) EBL absorption, for two fiducial redshifts ($z=0.5$ and 1) and for the parameters used in our EC-dominated model (see Fig.~\ref{fig:spectra_ec} and Table~\ref{tab:parameters}). For the attenuation of VHE photons, we considered several EBL models, as noted in the inset legend. The attenuated flux is compared against the 0.5~hr differential sensitivity\footnote{To obtain the 0.5~hr sensitivity curves of MAGIC and VERITAS, we scaled the publicly available  curves for 50~hr, respectively,  assuming that the sensitivity increases as $T^{-1/2}$, where $T$ is the observation time.} curves of the next-generation Cherenkov telescope array, i.e. CTA South \citep{Hassan2017} and two currently operating VHE telescopes, namely VERITAS and MAGIC. For a burst located at $z=0.5$, the EBL affects the spectrum already at energies $\gtrsim 100$~GeV, while the photon-photon absorption optical depth rises rapidly between $100$~GeV and $1$~TeV. High-quality spectra in this energy range can be used, in principle, to differentiate between EBL models, as shown in the top panel of Fig.~\ref{fig:spectrum_ebl}. For $z=1$, the flux at $\sim 1$~TeV is strongly attenuated for all the EBL models we considered. Still, CTA will be sensitive enough to detect emission up to $\sim 300$~GeV from that burst for almost all EBL models considered here.

\begin{figure*}
    \centering
    \includegraphics[width=2\columnwidth]{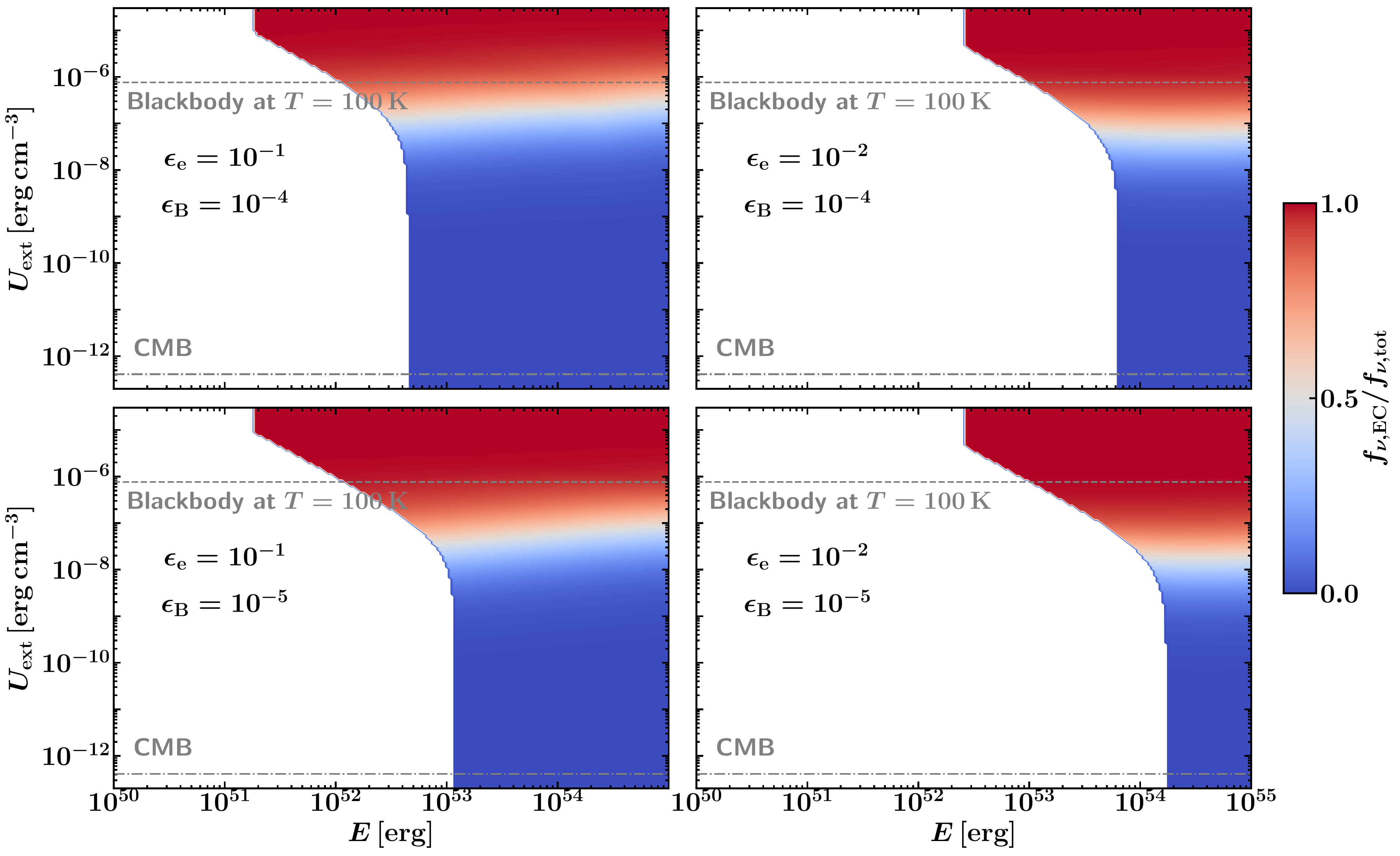}
    \caption{Detectability of the combined Compton (SSC + EC) signal from GRB afterglows at $100\,{\rm GeV}$ with CTA (assuming that the observation starts 0.5 hours after trigger and lasts for 0.5 hours), for different isotropic energies $E$ and energy densities of the external photon field $U_{\rm ext}$. Different panels show results for different combinations of $\epsilon_{\rm e}$ and $\epsilon_{\rm B}$ that are marked on each plot. The coloured area marks the parameter space of detectable afterglows (i.e. whose time-averaged 100~GeV flux over 0.5~hr is larger than the 0.5-hr CTA-South sensitivity). We also take into account the EBL attenuation and adopt the EBL model of \citet{Finke2010} (see dashed, black lines in Fig.~\ref{fig:spectrum_ebl}). The colour indicates the ratio of EC to the total Compton time-averaged fluxes, with red (blue) denoting EC-dominated (SSC-dominated) cases. In all panels, the horizontal lines indicate the energy density of the CMB (dashed-dotted) and of a black body with $T=100$~K (dashed). Other parameters used here are: $p=2.2$, $n=0.1\,{\rm cm^{-3}}$, and luminosity distance $d_{\rm L}=10^{28}\,{\rm cm}$.
    }
    \label{fig:cta_detectability}
\end{figure*}

We next discuss the detactibility of the combined Compton (SSC and  EC) signal at $100$~GeV by CTA, for a fiducial burst located at $z=0.5$ and different model parameters (e.g. $E$, $\epsilon_{\rm e}$, and $\epsilon_{\rm B}$). Using eqn.~(A5) from \citet{Sari2001} and eqn.~(\ref{eqn:slowspectrum}), we calculate the average Compton flux at 100 GeV\footnote{The EBL attenuation is taken into account. Here, we used the EBL model of \cite{Finke2010}.} over an interval of $T=0.5$~hr starting from $t=0.5$~hr, namely $\langle F_{\rm C}\rangle = T^{-1}\int_{t}^{t+T}dt'\left[F_{\rm SSC}(t') + F_{\rm EC}(t')\right]$, and compare it against the 0.5~hr CTA-South sensitivity at $100$~GeV (see Fig.~\ref{fig:spectrum_ebl}). We define a burst as {\sl detectable}, if $\langle F_{\rm C}\rangle$ exceeds the 0.5~hr CTA sensitivity. Our results are presented in Fig.~\ref{fig:cta_detectability}.

In all panels, the coloured regions indicate the parameter space of detectable bursts and the colour denotes the contribution of EC (red) and SSC (blue) to the total observed 100 GeV flux. Different panels show results for different combinations of the microphysical parameters $\epsilon_{\rm e}$ and $\epsilon_{\rm B}$. When EC makes only a small fraction of the total flux, we find that only rather powerful blasts may be detectable through their afterglow emission at high energies. For example, $E\gtrsim 5\times10^{53}$~erg is required for an SSC dominated GRB at $z=0.5$ to be detectable by CTA at 0.5~hr after the trigger (see upper right panel of Fig.~\ref{fig:cta_detectability}). However, when a dense ambient radiation field is present in the vicinity of a GRB, EC can significantly increase the production of $100~{\rm GeV}-1~{\rm TeV}$ photons. As a result, the detectability requirements on the blast isotropic equivalent energy are greatly reduced. This is illustrated by the extension of the red-coloured region towards lower $E$ values, if $U_{\rm ext}>10^{-8}-10^{-7}~{\rm erg \, cm^{-3}}$. Especially for $\epsilon_{\rm e}=0.1$ and $\epsilon_{\rm B}=10^{-5}$ (see lower left panel of Fig.~\ref{fig:cta_detectability}), the lower limit of $E$ is reduced by two orders of magnitude when $U_{\rm ext}$ increases from $\sim 10^{-8}$~erg cm$^{-3}$ to $10^{-5}~{\rm erg \, cm^{-3}}$. The region of the parameter space lying above the dashed horizontal line is unrealistic, as it implies energy densities exceeding that of a black-body photon field with temperature $T=100$~K, i.e. $U_{\rm ext}\simeq7.5\times 10^{-7} ~{\rm erg \, cm^{-3}}$. The typical value for $U_{\rm ext}$ can be several orders of magnitude below that of a black body. For instance, for ultraluminous infrared starburst galaxies (e.g., Arp 220), the energy density of external IR photon fields can be as large as $10^{-6} ~{\rm erg \, cm^{-3}}$ near the nucleus while for other star-forming galaxies (e.g., M82), $U_{\rm ext}$ can be about $10^{-10}$ to $10^{-9} ~{\rm erg \, cm^{-3}}$~\citep{wilson2014,scoville2015,Yoast-Hull2015,Perley2017,Yoast-Hull2017}. However, estimates of $U_{\rm ext}$ can vary as the size of the emitting region can be difficult to measure.

The parameter space of detectable events is also strongly dependent upon $\epsilon_{\rm e}$. The typical range for $\epsilon_{\rm e}$ values as obtained from afterglow modelling of the synchrotron component in GRBs is from $5\times10^{-3}$ to $0.3$~\citep{Cenko2011, Beniamini2017}. A larger value of $\epsilon_{\rm e}$ suggests that more of the shock energy is transferred into relativistic electrons, therefore producing more powerful Compton emission (either via SSC or EC). This, in turn, relaxes the requirements on the blast wave energy. The fraction of shocked fluid energy carried by the magnetic field, $\epsilon_{\rm B}$, affects only the detectability of SSC-dominated bursts (e.g. compare the top left and bottom left panels in  Fig.~\ref{fig:cta_detectability}). The value of $\epsilon_{\rm B}$ remains uncertain and may vary widely: $10^{-7}\sim10^{-1}$~\citep{zhang2015, Beniamini2016, Burgess2016}. With all other parameters fixed, a larger value of $\epsilon_{\rm B}$ increases the density of synchrotron photons that serve as targets for Compton scattering and, as a result, the SSC flux (see, e.g., eqn.~\ref{eqn:SSC_flux_scaling}). Thus, a smaller value of $\epsilon_{\rm B}$ indicates weaker SSC emissions, which will strengthen the requirements for a larger value of $E$ for VHE photons to be detected. This can be seen when transitioning from the top to bottom panels in Fig.~\ref{fig:cta_detectability}. 

\section{MAGIC Detection of GRB 190114C}
\label{sec:magic_grb_detection}

GRB 190114C \citep[at redshift $z=0.42$,][]{Selsing2019} is the first gamma-ray burst detected at sub-TeV energies by the MAGIC Cherenkov telescope \citep{Mirzoyan2019}. After the Swift-BAT trigger, the MAGIC detector showed a significance $>20 \, \sigma$ in the first 20 minutes of observations for energies $>300$~GeV. This VHE emission extended to $>300$~GeV provides a unique opportunity to test existing GRB afterglow models. 

Several studies aiming to interpret the VHE of GRB~190114C have already been presented. \citet{Ravasio2019}, for instance, argue that the afterglow emission at energies between 10~keV and 30~GeV should be produced by a single mechanism, either synchrotron or inverse Compton. Others propose that the SSC emission of GRB 190114C dominates over the synchrotron component at GeV energies \citep[e.g.][]{Fraija2019, Wang2019}.  \citet{Derishev2019} also showed that the sub-TeV emission of GRB~190114C can be SSC radiation produced at the early afterglow stage. In this section, we demonstrate that synchrotron radiation can explain the sub-GeV/GeV emission while radiation with energy beyond 100~GeV exceeds the synchrotron limit hence can only be explained by inverse Compton scattering. We also estimate the upper limit on the energy density of a putative ambient photon field using the LAT measurement at $\sim10^4$~s after the trigger and the MAGIC data.

\begin{figure}
\includegraphics[width=\columnwidth]{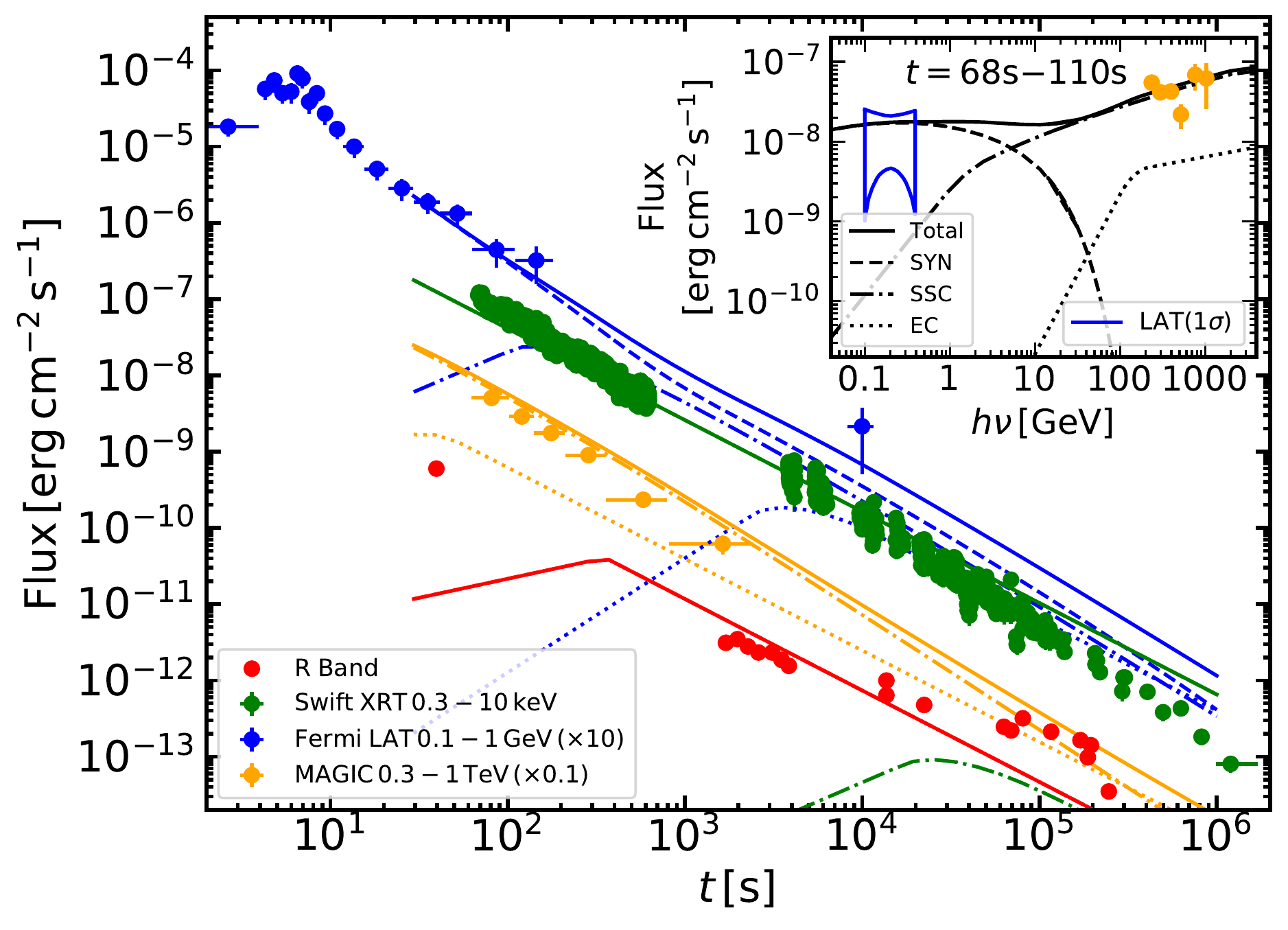}
\caption{Modeling of the afterglow light curves of GRB 190114C. The optical data are taken from \citet{Laskar2019}, the X-ray data are retrieved from Swift-XRT GRB light curve repository, the LAT data are taken from \citet{Ajello2019} and the MAGIC data are from \citet{2019MAGIC}. The model flux for the optical $R$ band has been modified due to the extinction by the host galaxy (assuming $A_V = 3.0$~mag). The MAGIC data has been corrected for EBL absorption and demonstrates the intrinsic light curve. The model-predicted light curves are displayed at times after the end of the coasting phase ($t\simeq30$~s), where we assumed that the initial bulk Lorentz factor is  $\Gamma_0=400$. In the inset plot we show the average spectrum from $68$~s to $110$~s. The yellow points show the VHE flux observed by MAGIC after EBL correction using the model of \citet{Dominguez2011}. The bowtie shows the $1\sigma$ contour of the power-law model fitted to the \textit{Fermi}-LAT data \citep{Ajello2019}, while different types of lines show the model spectra from various processes (for details, see inset legend). The parameters used here are: $E=5\times 10^{54}$~erg, $n=0.1$~cm$^{-3}$, $\epsilon_{\rm e}=0.05$, $\epsilon_{\rm B}=5\times 10^{-6}$, $p=2.6$, $\nu_0=0.02$~eV, $U_{\rm ext}=2.5\times 10^{-9}\,{\rm erg\,cm^{-3}}$. A coloured version of this plot is available online.}
\label{fig:grb190114c}
\end{figure}

Fig.~\ref{fig:grb190114c} shows the optical \citep{Laskar2019}, \textit{Swift}-XRT X-ray\footnote{\url{https://www.swift.ac.uk/xrt_curves/00883832/}}, \fermi-LAT gamma-ray \citep{Ajello2019}, and MAGIC VHE observations together with the optical, X-ray and gamma-ray light curves of GRB~190114C (colored lines) as obtained from our analytical model described in Sec.~\ref{sec:model} (for the parameters used, see figure caption).  As we are not considering the coasting phase of the blast wave in our model, we only show results for times larger than the deceleration time $t_{\rm dec}\simeq [E/(\pi n m_{\rm p} c^5 \Gamma_0^8)]^{1/3}(1+z)$, where $\Gamma_0$ is the initial bulk Lorentz factor. Here, we adopt $\Gamma_0=400$, which results in $t_{\rm dec}\simeq 30$~s.

In order to compare the effects of synchrotron and inverse Compton scattering on the electrons cooling, we estimate the values of $x$ and $y$ (for details, see Sec.~\ref{sec:model-1}). For this choice of parameters, $x$ decreases with time from $x\simeq 10$ at $t=50$~s to $x\simeq1$ at $10^6$~s while $y$ remains constant ($y \sim 0.2$). This indicates that SSC will dominate the cooling during most of the blast wave's deceleration phase ($t\lesssim 10^{5}$~s). 

The optical and X-ray fluxes consist mainly of synchrotron emission at all times. Given the adopted parameters, we find that $\nu_{\rm min}$ (given by eqn.~\ref{eq:synmin}) decreases from $20$~eV at $50$~s to $6\times10^{-6}~$eV at $10^{6}$~s. For $t\lesssim 200$~s, the peak of the synchrotron spectrum (in $F_\nu$ units) lies beyond the $R$ band (i.e. $\nu_{\rm min}>\nu_{\rm R}$), while at $t\gtrsim 200$~s we find $\nu_{\rm min} < \nu_{\rm R}$. The crossing of $\nu_{\min}$ through the $R$ band causes a break at $\sim 200$~s in the optical light curve. Note, however, that the model falls short in explaining the observed optical flux at $t\sim 30$~s. The bright early time optical emission might be produced by the reverse shock, not considered here \citep[see, e.g.][]{Laskar2019}. We also estimate the cooling break of the synchrotron spectrum using eqn.~(\ref{eqn:syncoolingbreak}) and find that $\nu_{\rm c}$ decreases only slightly from $5$~keV at $50$~s to $0.7~$keV at $10^{6}$~s. This indicates that the synchrotron cooling break $\nu_{\rm c}$ lies within the X-ray band. Our calculation shows the X-ray light curve decays as $t^{-\alpha}$ with $\alpha\sim 1.2$. This is consistent with the observed light curve. When electron cooling is dominated by SSC, as is the case here when $t\lesssim 10^5$~s, then the observed decay rate of the X-ray flux can only be explained by the propagation of a blast wave in a constant density medium \citep[see eqns.~B9 \&~C6 in][]{Panaitescu2000}. In contrast, if electron cooling was synchrotron-dominated, then  both the constant and the wind-like density profiles would lead to similar temporal decay rates \citep{Panaitescu2000,Ajello2019}.

The \fermi-LAT gamma-ray flux in the $0.1-1$~GeV energy range is dominated by synchrotron radiation (dashed blue line). At early times, the gamma-ray light curve, similar to the X-ray and optical ones, can be explained by synchrotron emission of electrons accelerated at the external shock wave. However, different from optical and X-ray emission, gamma-ray emitting electrons cannot cool efficiently through inverse Compton scattering due to KN suppression \citep[for more details, see][]{Nakar2009, Beniamini2015}. For instance, in Fig.~\ref{fig:grb190114c}, electrons with Lorentz factor $\gamma'\gtrsim 5\times 10^6$ at $t\sim 60$~s, which radiate synchrotron above $10$~MeV, can hardly cool via inverse Compton scattering. We therefore correct the synchrotron spectrum following \citet{Nakar2009} and the discussions in  Sec.~\ref{sec:model}.

It has also been suggested that the GeV emission could originate from inverse Compton scatterings \citep{Fraija2019}. However, neither EC nor SSC is likely to be the process powering the GeV emission of this burst, as we explain below. If EC dominated the GeV afterglow emission, this would require a small value of $\epsilon_{\rm B}$ (see eqn.~\ref{eqn:criticalcondition}): $\epsilon_{\rm B, -6}\lesssim 0.3n_0^{-2}\epsilon_{\rm e,-1}^{-1}U_{\rm ext, -6}^{-1}$. With such a small value of $\epsilon_{\rm B}$, it is difficult to simultaneously explain the flux in the X-ray and sub-TeV bands. Alternatively, the LAT flux could be attributed to SSC afterglow emission. However, it is difficult to make the SSC emission within the LAT energy band peak at times as early as $\sim10$~s for typical parameter values, as synchrotron photons at these times are typically up-scattered by electrons into the sub-TeV or TeV bands, and the light curve would rise instead of decay under this condition. For example, the peak of SSC can be estimated as $2\gamma_{\rm c}^2 \nu_{\rm c}\sim 50$~TeV at $t=50$~s for this particular case. For these reasons, synchrotron radiation is the most likely mechanism for producing the sub-GeV and GeV afterglow emission (see dashed blue line in Fig.~\ref{fig:grb190114c}).  

Although the emission in the LAT energy band is dominated by synchrotron radiation, the late-time measurement of the LAT flux (i.e. at $\sim 10^4$~s) is  crucial for constraining the parameters related to the inverse Compton scattering process, namely $n$ and $U_{\rm ext}$. The \fermi-LAT light curve from $t\gtrsim 10^2$~s to $10^4$~s can be described by a single power law. In our model, the SSC component in the LAT energy band rises at $\sim100$~s while the EC component rises at $\sim1000$~s. Both occur between the two \fermi-LAT data points. Given that the synchrotron, SSC, and EC light curves show similar temporal power-law decays (see blue curves in Fig.~\ref{fig:grb190114c}), neither the SSC nor the EC light curves at their peak can be brighter than the synchrotron flux at that time. Hence, we can calculate the maximum energy density of the external field $U_{\rm ext, max}\simeq3\times 10^{-9}\,  {\rm erg\,cm^{-3}}$.

SSC emission can also help in constraining the number density of the circumburst medium $n$. Assuming that SSC dominates the electron cooling, the synchrotron flux\footnote{Substituting $\nu_{\rm c}$ in eqn.~7 from \citet{Sari1998} with the value calculated by eqn.~\ref{eqn:gamma_c_2}.} $F_{\nu}^{\rm syn}\propto n^{(2-p)/(16-4p)}$ or $\propto n^{-0.1}$ for $p=2.6$. The SSC flux is written as $F_{\nu}^{\rm ssc}\propto n^{(p^2-14p+24)/(32-8p)}$ or $\propto n^{-0.5}$ for $p=2.6$ (see eqn.~\ref{eqn:SSC_flux_scaling}). The SSC flux dominating the \fermi-LAT at $t>500$~s is more sensitive to $n$, whereas the synchrotron flux which explains the X-ray and early \fermi-LAT emission is almost independent of $n$. As a result, the observed \fermi-LAT flux at $\sim 10^4$s provides constraint on $n$, which can be estimated as $n_{\rm max}\simeq0.1$~$\rm cm^{-3}$.

In Fig.~\ref{fig:grb190114c}, we also show our model applied to the VHE light curve (orange lines) and the time-averaged spectrum at time interval $68-110$~s. The MAGIC sub-TeV flux can be mostly explained as a result of inverse Compton scattering. The reason is that the photons' energy is much greater than the cutoff energy of synchrotron emission; the latter been $\sim 10$~GeV at $\sim 10^2$~s. The detection of high energy photons by MAGIC also helps to understand the underlying mechanism of this GRB and test existing EBL models.

Our modeling of GRB~190114C suggests that the EC flux can be similar to the SSC flux in the sub-TeV/TeV bands, especially at late time (i.e. $>$ several hours, see Fig.~\ref{fig:grb190114c}). Given the similarities in the EC and SSC emissions in sub-TeV/TeV energies, it may be difficult to distinguish between the two processes using only VHE spectra and light curves. One possible way out of this could lie in the synchrotron spectrum. Comparing the two illustrative examples in Fig.~\ref{fig:spectra}, we find that the KN correction makes the synchrotron spectrum of EC-dominated cases to appear harder than in SSC-dominated ones. Compared with a SSC-dominated case, a harder synchrotron spectrum for an EC-dominated one is expected at a frequency of $h\nu=\gamma'^{*2}\Gamma h e B'/2\pi m_{\rm e} c\approx30 \sqrt{n_0\epsilon_{\rm B,-4}}\,{\rm keV}$. Therefore, observations in hard X-rays, i.e., between the  \fermi-GBM and \fermi-LAT bands, could help us further constrain the relative contributions of EC and SSC emissions.

Here, we discussed the synchrotron and inverse Compton emission from a forward shock propagating into a constant density circumstellar environment, but it is also possible that a wind-like density profile can explain the afterglow emissions. \citet{Ajello2019} showed that the synchrotron model in a wind-like circumstellar environment works well in explaining the X-ray and sub-GeV/GeV gamma-ray afterglow light curves. However, the authors assumed that electrons are cooling mainly via synchrotron radiation, while inverse Compton cooling was neglected, which may not be a valid assumption, especially at later times, when both EC and SSC are in Thomson regime and electrons can cool via inverse Compton scattering.
A detailed study of the multi-wavelength afterglow emission for a wind-like density profile could be the topic of a future publication, following the release of the MAGIC data.

\section{Summary and Discussion}
\label{sec:discussion}
In this paper, we perform a systematic study of the Compton emission in GRB afterglows, with the inclusion of a narrow-band ambient radiation field as a source of scattering. We calculated synchrotron, SSC, and EC spectra and light curves produced by a power-law distribution of electrons accelerated at the relativistic shock during its deceleration phase, as it sweeps up matter from a constant-density circumburst medium. Similar to the synchrotron radiation, we find that the flux at the peak of EC remains constant in both slow and fast cooling regimes for adiabatic hydrodynamic evolution of the blast wave, while the peak of the SSC component decreases with time.

The calculations of inverse Compton scattering indicate that either EC or SSC can explain the high energy emission at energies beyond $100$~MeV. We find that SSC may dominate the cooling of electrons over EC, except when there is a dense ambient IR radiation (as observed in some star-forming galaxies) or a low-density circumburst medium (see eqn.~\ref{eqn:criticalcondition}). 

We also discuss the detectability of VHE afterglow emission by existing and future gamma-ray instruments when the EBL attenuation is considered. When a dense ambient radiation field is present in the vicinity of a GRB, EC scattering can significantly increase the emission of 100~GeV--1~TeV photons. As a result, the detectability requirements on the blast isotropic equivalent energy are greatly reduced. Being about one order of magnitude more sensitive than current Cherenkov telescopes, CTA should be capable of detecting sub-TeV and TeV photons with flux as low as $\nu F_{\nu}\sim 10^{-12}$~${\rm erg\,cm^{-2}\,s^{-1}}$ (with an observation time 0.5~h). This also means that a burst may be detectable with CTA even at very late times, assuming a power-law decay of the flux $\propto t^{(9p^2 - 38p + 24)/(32 - 8p)}$ for SSC-dominated cases or $\propto t^{-(3p-2)/4}$ for EC-dominated ones. In the CTA era, we expect more detections of GRB afterglows in GeV and TeV bands in host galaxies with regions of dense IR radiation. 

We apply our analytical afterglow emission model to the GRB~190114C, the first gamma-ray burst detected at sub-TeV energies by the MAGIC Cherenkov telescope. We find that the optical and X-ray light curves can be explained by synchrotron emission of particles accelerated in a power-law energy spectrum with slope $p=2.6$ at a relativistic adiabatic blast wave of energy $E\simeq 5\times10^{54}$~erg propagating in a circumburst medium with density $n=0.1$~cm$^{-3}$. We also find that the \fermi-LAT light curve is synchrotron dominated. The \fermi-LAT measurement at $10^4$~s after trigger is crucial for setting an upper limit on the energy density of a putative IR photon field (i.e. $U_{\rm ext}\lesssim 3\times 10^{-9}\,  {\rm erg\,cm^{-3}}$). Studying the spectrum at $68-110$~s, we find the \fermi-LAT flux at $100$~MeV is comparable to the MAGIC VHE flux at $100$~GeV. It gives a strong support that the VHE emission is produced by inverse Compton scattering while the sub-GeV emission originates from synchrotron. We also show that the observed VHE flux decays as $t^{-1.4}$, which fits well with our model. %and it is dominated by SSC instead of EC.

\section*{Acknowledgements}
The authors thank the anonymous referees for their constructive comments that helped to improve the manuscript. IMC and MP acknowledges support from the Fermi Guest Investigation grant 80NSSC18K1745. MP also acknowledges support from the Lyman Jr.~Spitzer Postdoctoral Fellowship. JMRB acknowledges the support from the Mexican Council of Science and Technology (CONACYT) for the Postdoctoral Fellowship under the program Postdoctoral Stays Abroad. DG acknowledges support from the NASA ATP NNX17AG21G, the NSF AST-1910451 and the NSF AST-1816136 grants.
% The Acknowledgements section is not numbered. Here you can thank helpful
% colleagues, acknowledge funding agencies, telescopes and facilities used etc.

%%%%%%%%%%%%%%%%%%%%%%%%%%%%%%%%%%%%%%%%%%%%%%%%%%



\bibliographystyle{mnras}
\bibliography{bib} % if your bibtex file is called example.bib

%%%%%%%%%%%%%%%%%%%%%%%%%%%%%%%%%%%%%%%%%%%%%%%%%%

%%%%%%%%%%%%%%%%% APPENDICES %%%%%%%%%%%%%%%%%%%%%

\appendix

\section{External Compton Scattering Spectra and Light Curves}
\label{appndx:ec}
Here, we derive analytical expressions for the high-energy photon spectra and light curves produced by external Compton scattering in the fast and slow cooling regimes.

The average frequency of Thomson scattered photons in the shocked fluid frame is  $\nu^{\prime \rm EC} \approx \frac{4}{3}\gamma^{\prime 2} \Gamma \nu_0$, where $\nu_0$ is the frequency of external photons (and $\epsilon_0=h\nu_0$), as measured in the observer frame. The peak spectral power can be estimated as
\begin{equation}
\label{eqn:P_nu_max}
P^{\prime \rm EC}_{\nu',\rm max}\approx\frac{P'_{\rm EC}}{\nu^{\prime \rm EC}}=\frac{\sigma_{\mathrm{T}} c \left(\Gamma U_{\rm ext}\right)}{\nu_0},
\end{equation}
which depends solely on the properties of the external photon field, as long as the scattering occurs in the Thomson limit
\begin{equation}
\label{eqn:gamma_th}
\gamma' \lesssim \gamma'^* \equiv \frac{m_{\mathrm{e}}c^2}{\Gamma \epsilon_0} = 1.5 \times 10^5 \left(\frac{10^2}{\Gamma} \right)  \left( \frac{0.1 \,{\rm eV}}{\epsilon_0}\right).
\end{equation}
Henceforth, we consider only scatterings in the Thomson regime. In the observer frame, the average energy of observed photons after scattering is approximately: 
\begin{equation}
\label{eqn:nu_obs}
\nu^{\rm EC}(\gamma') \approx \frac{4}{3}\Gamma^2 \gamma'^{ 2} \nu_0.
\end{equation}
In order to obtain the observed net spectrum $F^{\rm EC}_{\nu}$, we need to integrate the spectrum of a single scattering over all electrons. The accelerated electron distribution (see eqn.~\ref{eqn:injection_rate}) is modified by the radiative cooling and can be written as
\begin{equation}
\label{eqn:edistri_fast}
N(\gamma')\propto\left
\{\begin{array}{ll}
\gamma'^{-2} &\gamma'_{\mathrm{c}}<\gamma'<\gamma'_{\mathrm{min}}\\
\gamma'^{-p-1} &\gamma'>\gamma'_{\mathrm{min}}
\end{array}
\right.,
\end{equation}
in the fast cooling regime (i.e. $\gamma'_{\rm min}>\gamma'_{\rm c}$) or
\begin{equation}
\label{eqn:edistri_slow}
N(\gamma')\propto\left\{
\begin{array}{ll}
\gamma'^{-p} &\gamma'_{\mathrm{min}}<\gamma'<\gamma'_{\mathrm{c}}\\
\gamma'^{-p-1} &\gamma'>\gamma'_{\mathrm{c}}
\end{array}
\right.,
\end{equation}
in the slow cooling regime (i.e. $\gamma'_{\rm min}<\gamma'_{\rm c}$). Here, we introduce two characteristic frequencies that will prove useful for later: $\nu^{\rm EC}_{\rm min} \equiv \nu^{\rm EC}(\gamma'_{\rm min})$, $\nu^{\rm EC}_{\rm c} \equiv \nu^{\rm EC}(\gamma'_{\rm c})$, determined using eqn.~(\ref{eqn:nu_obs}). The low-energy part of the net spectrum (i.e. $\nu<\min[\nu^{\rm EC}_{\rm c}, \nu^{\rm EC}_{\min}]$) is the sum of the low-energy tails of the single-particle Compton spectrum from all electrons, and as such $F^{\rm EC}_{\nu} \propto \nu$. The remaining part of the spectrum can be calculated according to the relationship
\begin{equation}
F^{\rm EC}_{\nu} \rm{d}\nu \propto P_{\rm EC}(\gamma')[N(\gamma') \rm{d}\gamma'],
\end{equation}
where $P_{\rm EC}(\gamma') \simeq\Gamma^2P'_{\rm EC}(\gamma')\propto\gamma'^2$ is the EC power in the observer frame and is determined using eqn.~(\ref{eqn:EC_power_single_particle}).

The total spectrum in the slow cooling regime can be written as 
\begin{equation}
\label{eqn:slowspectrum}
\resizebox{\linewidth}{!}{$
F^{\rm EC}_{\nu}= F^{\rm EC}_{\nu,\mathrm{max}} \times\left\{
\begin{array}{ll}
(\nu/\nu^{\rm EC}_{\mathrm{min}}) & \nu<\nu^{\rm EC}_{\mathrm{min}} \\ \\
(\nu/\nu^{\rm EC}_{\mathrm{min}})^{-(p-1)/2} & \nu^{\rm EC}_{\mathrm{min}}<\nu<\nu^{\rm EC}_{\mathrm{c}} \\ \\
(\nu^{\rm EC}_{\mathrm{c}}/\nu^{\rm EC}_{\mathrm{min}})^{-(p-1)/2} \, (\nu/\nu^{\rm EC}_{\mathrm{c}})^{-p/2} & \nu>\nu^{\rm EC}_{\mathrm{c}}
\end{array}
\right. ,$}
\end{equation}
while in the fast cooling regime it is given by
\begin{equation}
\label{eqn:fastspectrum}
\resizebox{\linewidth}{!}{$
F^{\rm EC}_{\nu}= F^{\rm EC}_{\nu,\mathrm{max}}\times\left\{
\begin{array}{ll}
(\nu/\nu^{\rm EC}_{\mathrm{c}}) & \nu<\nu^{\rm EC}_{\mathrm{c}} \\ \\
(\nu/\nu^{\rm EC}_{\mathrm{c}})^{-1/2} & \nu^{\rm EC}_{\mathrm{c}}<\nu<\nu^{\rm EC}_{\mathrm{min}} \\ \\
(\nu^{\rm EC}_{\mathrm{min}}/\nu^{\rm EC}_{\mathrm{c}})^{-1/2} \, (\nu/\nu^{\rm EC}_{\mathrm{min}})^{-p/2}&         \nu>\nu^{\rm EC}_{\mathrm{min}}
\end{array}
\right.,$}
\end{equation}
where $F^{\rm EC}_{\nu,\rm max}$ is the observed peak flux
\begin{equation}
\label{eqn:peak_flux}
F^{\rm EC}_{\nu,\rm max}\equiv \frac{L^{\rm EC}_{\nu,\rm max}}{4\pi d_{\rm L}^2} (1+z).
\end{equation}
Here, $d_{\rm L}$ is luminosity distance of the source and $L^{\rm EC}_{\nu,\rm max}\equiv (4/3) \pi R^3  n \Gamma P^{\prime \rm EC}_{\nu',\rm max}$ is the maximum spectral luminosity. 

In the EC-dominated regime (for details, see Sec.~\ref{sec:model-1}), we obtain simple expressions for the peak flux, minimum, and cooling frequencies of the EC spectrum:
\begin{align}
\label{eqn:fmax_ec}
F^{\rm EC}_{\nu,\mathrm{max}} = 6.1\times 10^{-3} \epsilon_{\mathrm{0,eV}}^{-1} U_{\mathrm{ext},-6} E_{54} d_{\mathrm{L},28}^{-2} (1+z) \,  [\mathrm{nJy}],\\ 
\nu^{\rm EC}_{\mathrm{min}}= 0.64 \left(\frac{p-2}{p-1}\right)^{2} \epsilon_{\mathrm{e,-1}}^{2} \epsilon_{\mathrm{0,eV}} E_{54}^{1/2} n_0^{-1/2} t_{\mathrm{d}}^{-3/2} (1+z)^{1/2} \,[\mathrm{GeV}],\\
\nu^{\rm EC}_{\mathrm{c}} =1.1\times 10^4 \epsilon_{\mathrm{0,eV}} U_{\mathrm{ext},-6}^{-2} E_{54}^{-1/2} n_0^{1/2} t_{\mathrm{d}}^{-1/2} (1+z)^{-1/2} \, [\mathrm{GeV}],
\end{align}
where $\epsilon_{\mathrm{0,eV}}=\epsilon_{\mathrm{0}}/[1\,\mathrm{eV}]$ and
$t_{\mathrm{d}}$ is the time in the observer frame normalized to 1 day. 

We present next expressions for the temporal evolution of the EC flux, assuming $p=2.2$, in both cooling regimes. For the slow cooling regime, we find
\begin{align}
\label{eqn:slowspectrum1a}
\frac{F^{\rm EC}(t)_{\nu < \nu^{\rm EC}_{\mathrm{min}}}}{8.6\times10^2(1+z)^{0.5} \mathrm{[nJy]}}= \epsilon_{\mathrm{e,-1}}^{-2} U_{\mathrm{ext,-6}} E_{54}^{0.5} n_0^{0.5} t_{\mathrm{d}}^{1.5} \nu_{\mathrm{GeV}} d_{\mathrm{L,28}}^{-2}, \\
\label{eqn:slowspectrum1b}
\frac{F^{\rm EC}(t)_{\nu^{\rm EC}_{\mathrm{min}}<\nu<\nu^{\rm EC}_{\mathrm{c}}}}{ \mathrm{2.6\times10^{-3}(1+z)^{1.3}[nJy]}}= \epsilon_{\mathrm{e,-1}}^{1.2} U_{\mathrm{ext,-6}} E_{54}^{1.3} n_0^{-0.3} t_{\mathrm{d}}^{-0.9} \nu_{\mathrm{GeV}}^{-0.6} d_{\mathrm{L,28}}^{-2},  \\ 
\label{eqn:slowspectrum1c}
\frac{F^{\rm EC}(t)_{\nu>\nu^{\rm EC}_{\mathrm{c}}}}{ 8.2\times10^{-2}(1+z)^{1.05} \mathrm{[nJy]}}=\epsilon_{\mathrm{e,-1}}^{1.2} E_{54}^{1.05} n_0^{-0.05} t_{\mathrm{d}}^{-1.15} \nu_{\mathrm{ GeV}}^{-1.1} d_{\mathrm{L,28}}^{-2},
\end{align}
where $\nu_{\rm GeV}\equiv \nu/(2.4\times10^{23} \ {\rm Hz})$.
Accordingly, the expressions for the fast cooling regime are
\begin{align}
\label{eqn:fastspectrum1a}
\frac{F^{\rm EC}(t)_{\nu < \nu^{\rm EC}_{\mathrm{c}}}}{1.4\times10^{-3}(1+z)^{1.5} \mathrm{[nJy]}}=U_{\mathrm{ext,-6}}^3 E_{54}^{1.5} n_0^{-0.5} t_{\mathrm{d}}^{0.5}  \nu_{\mathrm{GeV}}d_{\mathrm{L,28}}^{-2}, \\
\label{eqn:fastspectrum1b}
\frac{F^{\rm EC}(t)_{\nu^{\rm EC}_{\mathrm{c}}<\nu<\nu^{\rm EC}_{\mathrm{min}}}}{4.5 (1+z)^{0.75} \mathrm{[nJy]}}=E_{54}^{0.75} n_0^{0.25} t_{\mathrm{d}}^{-0.25} \nu_{\mathrm{GeV}}^{-0.5} d_{\mathrm{L,28}}^{-2}, \\
\label{eqn:fastspectrum1c}
\frac{F^{\rm EC}(t)_{\nu>\nu^{\rm EC}_{\mathrm{min}}}}{8.2\times10^{-2}(1+z)^{1.05}}=\epsilon_{\mathrm{e,-1}}^{1.2} E_{54}^{1.05} n_0^{-0.05} t_{\mathrm{d}}^{-1.15} \nu_{\mathrm{ GeV}}^{-1.1} d_{\mathrm{L,28}}^{-2}.
\end{align}

\begin{figure*}
    \centering
    \begin{subfigure}{0.45\textwidth}
        \includegraphics[width=\textwidth]{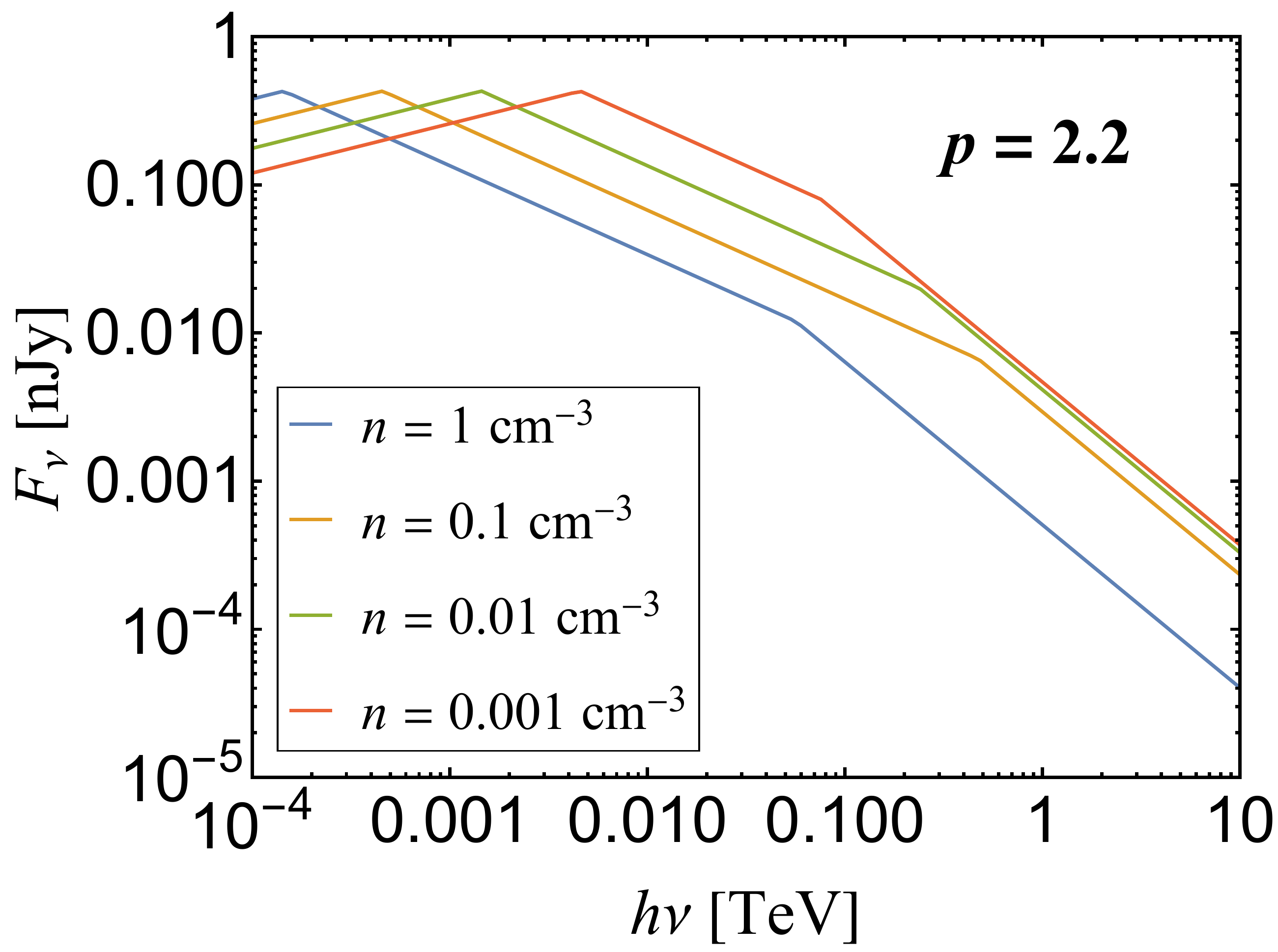}
        \label{fig:spectra_n}
    \end{subfigure}
    \begin{subfigure}{0.45\textwidth}
        \includegraphics[width=\textwidth]{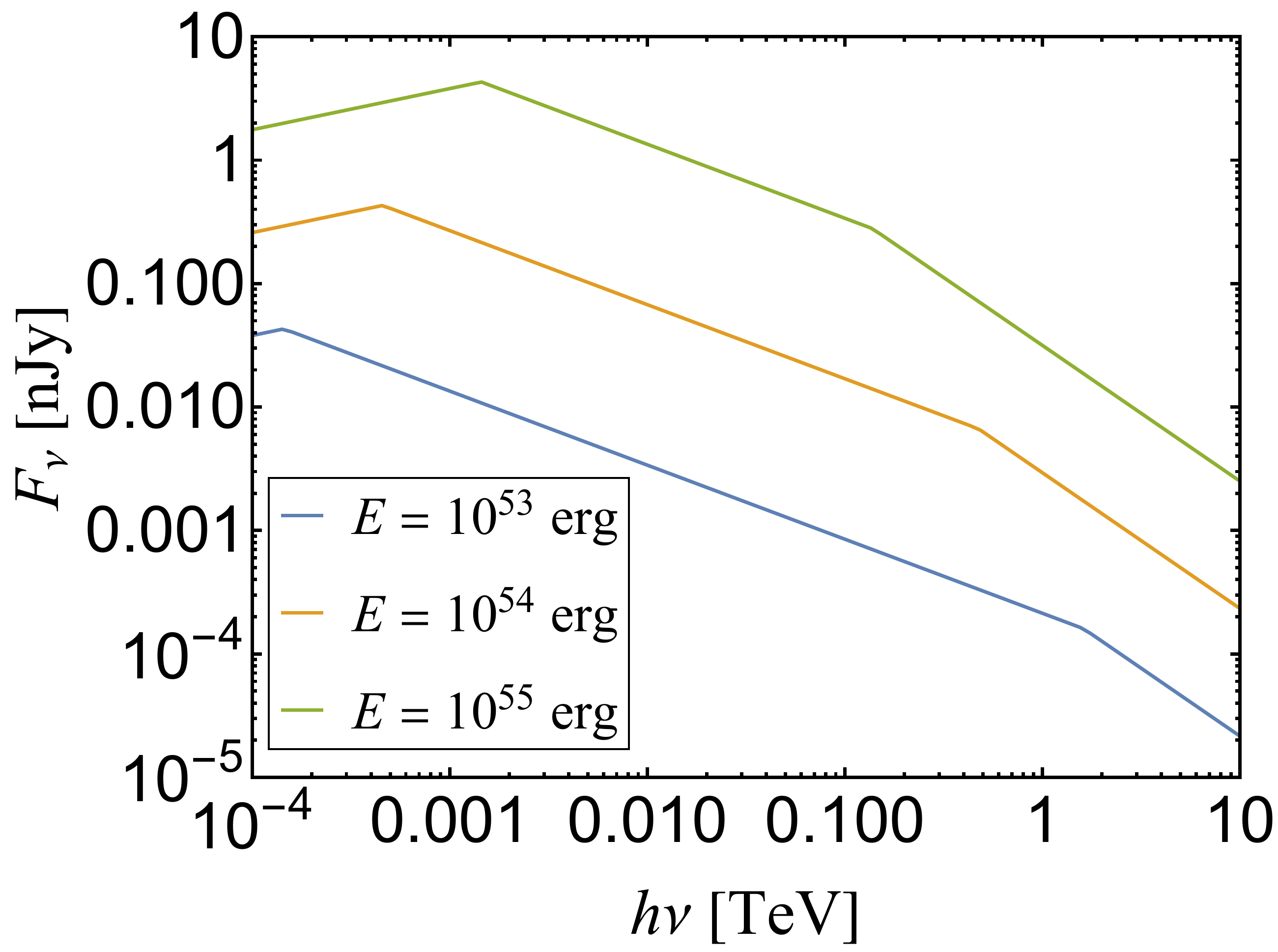}
        \label{fig:spectra_e}
    \end{subfigure}
    \begin{subfigure}{0.45\textwidth}
        \includegraphics[width=\textwidth]{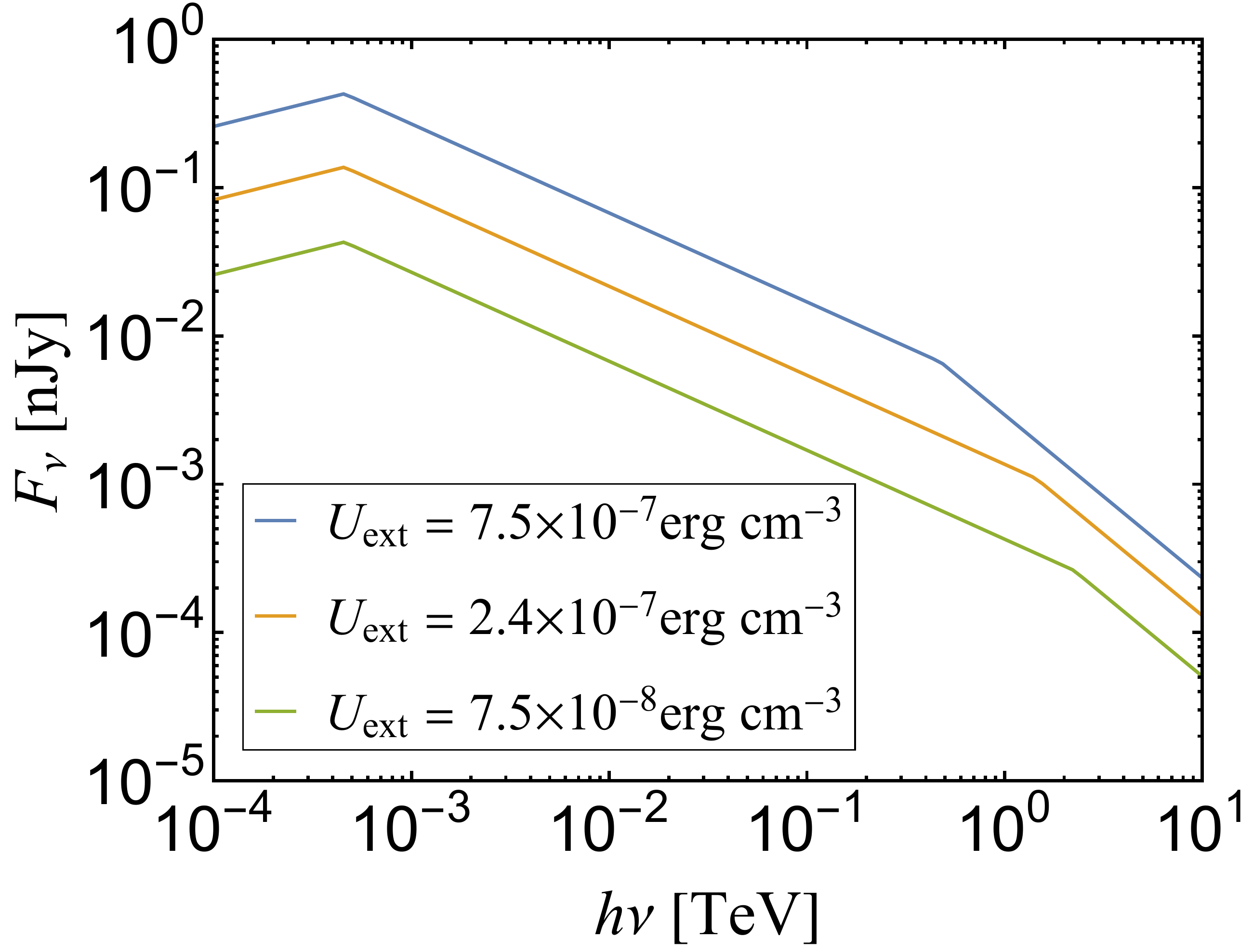}
        \label{fig:spectra_t}
    \end{subfigure}
    \begin{subfigure}{0.45\textwidth}
        \includegraphics[width=\textwidth]{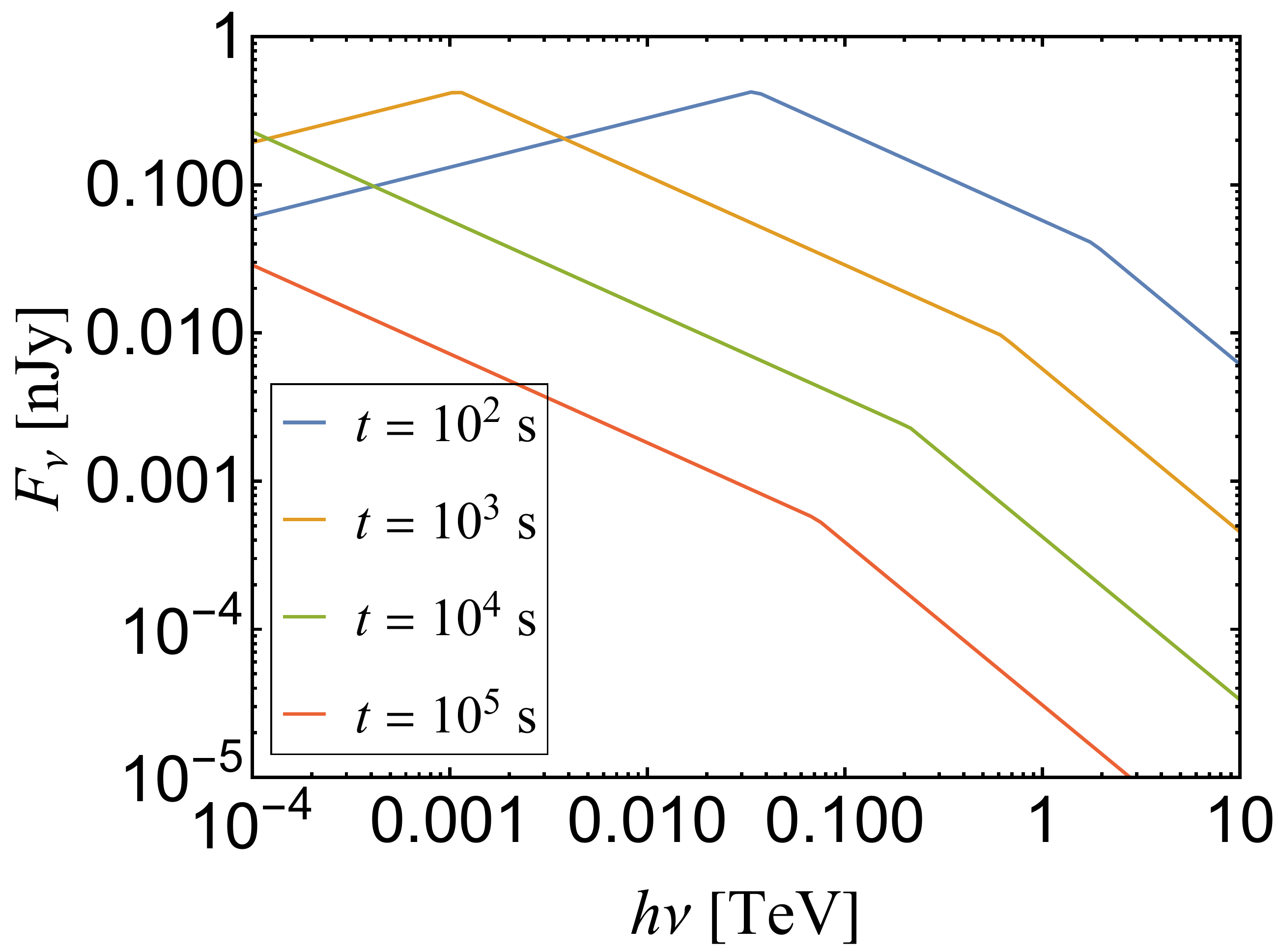}
        \label{fig:spectra_uext}
    \end{subfigure}
    \caption{External Compton spectra computed using eqns.~(\ref{eqn:slowspectrum1a})-(\ref{eqn:slowspectrum1c}) for different model parameters. From top left and in clockwise order we vary the number density of the circumburst medium $n$, the initial blast wave energy $E$, the observer time $t$, and the external photon field energy density $U_{\rm ext}$. The parameters used, unless otherwise specified, are: $p=2.2$, $n=0.1\,\mathrm{cm^{-3}}$, $E=10^{55}\,\mathrm{erg}$, $\epsilon_{\mathrm{e}}=0.1$, $\epsilon_0=0.02\,\mathrm{eV}$, $d=10^{28}\,\mathrm{cm}$, $z=0.5$, and $t=0.5\,\mathrm{h}$. A coloured version of this plot is available online.}
    \label{fig:spectra_net}
\end{figure*}

All expressions derived so far are valid for scatterings occurring in the Thomson limit (see eqn.~\ref{eqn:gamma_th}). Electrons with Lorentz factor greater than $\gamma'^*$ scatter  $h\nu_0$ photons into the KN regime, where the scattering cross section is proportional to $\sigma_{\rm T}\ln{2x}/x$ \citep{Blumenthal1970}; here, $x=\epsilon_0/m_{\mathrm{e}}c^2$. A direct effect of this is the suppression of high-energy photon production, which happens to the observed photons with energies above
\begin{equation}
\label{eqn:EC_KN_cutoff}
h\nu_{\mathrm{KN}}\sim \Gamma^2\gamma'^{*2}\epsilon_0=\frac{(m_{\mathrm{e}}c^2)^2}{\epsilon_0} \approx 3\,\mathrm{TeV} \left(\frac{0.1\mathrm{eV}}{\epsilon_0}\right).
\end{equation}
As long as $\nu < \nu_{\rm KN}$, one can safely use the analytical expressions for $F_\nu(t)$ presented here. 

Fig.~\ref{fig:spectra_net} shows EC spectra of GRB afterglows computed using eqns.~(\ref{eqn:slowspectrum})--(\ref{eqn:fastspectrum}) for different parameters. In the upper left panel, we fix all parameters except $n$ and compare spectra at a given time. A less dense ISM results in the shock taking longer time to slow down. So for the given time, the shock in the less dense ISM has a larger Lorentz factor, indicating a higher peak frequency $\nu^{\rm EC}_{\rm peak}$. We also notice that the break frequency of the EC spectrum $\nu^{\rm EC}_{\rm c}$ increases with $n$ when $n<1~{\rm cm^{-3}}$, but $\nu^{\rm EC}_{\rm c}$ shrinks significantly for $n\gtrsim1~{\rm cm^{-3}}$. This is because when $n>1~{\rm cm^{-3}}$, SSC starts to dominate the cooling of electrons. Therefore, the EC emission at $100$~GeV and above drops significantly. In EC-dominated cases, the flux of $100$~GeV to $1$~TeV photons depends on $n$ weakly, which may provide a method to estimate $E$ (see Eqn.~\ref{eqn:slowspectrum1c}). 

The upper right panel of Fig.~\ref{fig:spectra_net} demonstrates the dependence of flux on the isotropic energy $E$. It is noticed that a more energetic burst will not only produce a larger flux in all bands, but also increase the break frequency $\nu^{\rm EC}_{\rm min}$ and $\nu^{\rm EC}_{\rm c}$, since it can accelerate particles to higher energy.

The lower left panel of Fig.~\ref{fig:spectra_net} illustrates the influence of the energy density of the external field $U_{\rm ext}$ on the EC spectrum. We notice that a stronger ambient photon field will increase the EC emission and significantly enhance the VHE flux. In our model the hydrodynamics of the shock is independent of the external photon field, and $\nu^{\rm EC}_{\rm min}$ does not change for different values of $U_{\rm ext}$. But electrons cool faster due to a stronger photon field. Thus, the break frequency of EC spectrum, $\nu^{\rm EC}_{\rm c}$, drops as $U_{\rm ext}$ increases.

The lower right panel of Fig.~\ref{fig:spectra_net} shows the time-dependent EC spectra. As time evolves, the external shock gradually decelerates, and particles becomes less energetic. Hence, the peak frequency of scattered photons $\nu^{\rm EC}_{\rm peak}$ decreases. We find it interesting that the peak flux $F^{\rm EC}_{\rm peak}$ remains unchanged as time evolves.

\section{The Effect from Klein-Nishina Suppression of External Compton on Electron Cooling}\label{appndx:yKN}

The electron cooling can be significantly affected by the KN suppression, especially for electrons with large Lorentz factor. This would lead to $y\equiv U^\prime_{\rm ext}/U^\prime_{\rm B}$ being $\gamma'$ dependent which results in a strong signature on both spectra of synchrotron and inverse Compton. The dependence of $x\equiv U^\prime_{\rm syn}/U^\prime_{\rm B}$ (or Compton parameter for SSC) on $\gamma'$ has been fully studied by \citet[]{Nakar2009}; here we only show the dependence of $y$ (Compton parameter for EC) on $\gamma'$ using a similar method.

Under the assumption that the Lorentz factor of electron is much greater than the bulk motion (i.e. $\gamma'\gg \Gamma$) and the external photon field is a grey body, $y$ can be approximated as:
\begin{equation}\label{eqn:yKN0}
    y(\gamma)=\frac{1}{U'_{\mathrm{B}}}\int^{\infty}_{0}B'_{\nu'}\int^{1}_{-1}\frac{(1-\mu)\sigma_{\mathrm{KN}}[\frac{\nu'}{\tilde{\nu'}}(1-\mu)]}{1+\frac{\nu'}{\tilde{\nu'}}}\mathrm{d}\mu\mathrm{d}\nu',
\end{equation}
where $B'_{\nu'}$ is the energy density of the grey body in the comoving frame of the shock and it is proportional to $\nu'^3\left(e^{\frac{h\nu'}{\Gamma \epsilon_{0}}}-1\right)^{-1}$. Additionally, $\mu$ is the cosine of the angle between the upscattered photon and the momentum of the electron in the comoving frame of the shock,   $\sigma_{\mathrm{KN}}[x]$ is the KN cross section for scattering of photons with energy $h\nu'=x m_{\mathrm{e}} c^2$ in the electron’s rest frame, and $\tilde{\nu'}$ is the maximum energy in the comoving frame of photons that can be upscattered in the Thomson regime by an electron with Lorentz factor $\gamma'$ and $\gamma' \tilde{\nu'}=m_{\mathrm{e}}c^2$. The integral over $\mu$ yields $1$ in the Thomson regime ($\nu'/\tilde{\nu'}\ll 1$), while at the deep KN regime ($\nu'/\tilde{\nu'}\gg 1$) it becomes $\ln{2\nu'}/\nu'^2$. Therefore, eqn.~\ref{eqn:yKN0} can be written as:
\begin{equation*}
    y(\gamma')\approx\frac{1}{U'_{\mathrm{B}}}\int^{1}_{0}B'_{\nu'}\mathrm{d}\left(\frac{\nu'}{\tilde{\nu'}}\right)+\int^{\infty}_{1}\frac{3}{8}B'_{\nu'}\frac{\ln{\frac{2\nu'}{\tilde{\nu'}}}}{\frac{\nu'}{\tilde{\nu'}}}\mathrm{d}\left(\frac{\nu'}{\tilde{\nu'}}\right).
\end{equation*}
When $\tilde{\nu'}\gg \Gamma\epsilon_0$, the contribution from the second term is zero due to the exponential cut-off in $B'_{\nu'}$. When $\tilde{\nu'}\ll \Gamma\epsilon_0$, the second term will become dominant, and $y(\gamma')\propto\tilde{\nu'}^2\ln{\gamma'}$. Therefore, if we neglect logarithmic terms, we have:

\begin{equation}
\label{eqn:ykn}
y(\gamma')= \frac{\Gamma^2 U_{\mathrm{ph}}}{U'_{\mathrm{B}}} \times\left\{
\begin{array}{ll}
1 & \gamma'<\gamma'^* \\
\left(\frac{\gamma'}{\gamma'^*}\right)^{-2} & \gamma'>\gamma'^*
\end{array}
\right. ,
\end{equation}
where $\gamma'^*=\frac{m_{\rm e}c^2}{\Gamma\epsilon_0}$.

\section{The Wind-like Density Profile}
The radiation process only depends on the particle density near the shock front at the observation time. Thus, there is no difference in the instantaneous spectra computed for a homogeneous circumburst medium and a medium with wind-like number density $n=A r^{-s}$. However, since the dynamics of the blast wave will be different, the temporal evolution of the flux will be changed.

Eqns.~\ref{eqn:bw_radius} and \ref{eqn:bw_Gamma} now become
\begin{equation}
    R\simeq\left[\frac{Et}{m_{\rm p}cA(1+z)}\right]^{\frac{1}{4-s}},
\end{equation}
\begin{equation}
    \Gamma\simeq\left[\frac{E(1+z)^{3-s}}{m_{\rm p}c^{5-s}At^{3-s}}\right]^{\frac{1}{2(4-s)}}.
\end{equation}
For a wind ejected by the GRB progenitor at a constant speed (i.e, $s=2$), the expressions for the blast wave radius and Lorentz factor read
\begin{equation}
    R=\left(\frac{Et}{\pi m_{\rm p}cA(1+z)}\right)^{1/2},
\end{equation}
\begin{equation}
    \Gamma=\left(\frac{E(1+z)}{16\pi m_{\rm p}c^3At}\right)^{1/4}.
\end{equation}
The definitions of $x$ and $y$ are valid for any density profiles, yet we have to recalculate their value based on $\gamma'_{\rm min}$ and $\gamma'_{\rm c}$ in the wind case.

For the wind density profile, the value of $y$ is proportional to $t$ rather than constant as in the constant case
\begin{equation}
    y=0.4 \epsilon_{\rm B,-4}^{-1}A_{*}^{-2}U_{\rm ext,-6} E_{54} t_{\rm h}/(1+z),
\end{equation}
where $A_{*}= A/(3.0\times 10^{35} {\rm cm^{-1}})$ and $t_{\rm h}\equiv t/{\rm 1 \, hour}$. This suggests that EC plays a more important role at later time than SSC.
The temporal evolution of $x$ can be written as:
\begin{itemize}
\item SSC-dominated ($x\gg y>1$)
\begin{equation}
x\simeq\left\{
\begin{array}{ll}
\sqrt{\dfrac{\epsilon_{\mathrm{e}}}{\epsilon_{\mathrm{B}}}} &\gamma'_{\rm min} > \gamma'_{\rm c}\\
\left(\dfrac{\epsilon_{\mathrm{e}}}{\epsilon_{\mathrm{B}}}\right)^{\frac{1}{4-p}} \left(\dfrac{t}{t_0}\right)^{\frac{2-p}{4-p}} &\gamma'_{\rm min} < \gamma'_{\rm c}
\end{array}
\right.;
\end{equation}
\item EC-dominated ($y\gg x>1$)
\begin{equation}
x\simeq\left
\{\begin{array}{ll}
\dfrac{1}{y}\dfrac{\epsilon_{\mathrm{e}}}{\epsilon_{\mathrm{B}}} &\gamma'_{\rm min} > \gamma'_{\rm c}\\
\dfrac{1}{y^{3-p}}\dfrac{\epsilon_{\mathrm{e}}}{\epsilon_{\mathrm{B}}} \left(\dfrac{t}{t_0}\right)^{2-p} &\gamma'_{\rm min} < \gamma'_{\rm c}
\end{array}
\right.,
\end{equation}
\end{itemize} 
where $t_0=0.08\left(\frac{p-2}{p-1}\right)\epsilon_{\rm e,-1}\epsilon_{\rm B,-4}A_{*}(1+z)\,{\rm [h]}$.

Similar to Sec.~\ref{sec:model}, we derive the following critical condition 
\begin{equation}
    \epsilon_{\rm e,-1}^{1-p} \epsilon_{\rm B,-4}^{-1} A_{*}^{p-6} U_{\rm ext,-6}^{4-p} E_{54}^{4-p} \left(\frac{t_{\rm h}}{1+z}\right)^2 \approx 200\times30^{3-p}\left(\frac{p-2}{p-1}\right)^{p-2}.
\end{equation}
If the left-hand side is smaller than the right-hand side of the equation above, SSC dominates electron cooling over EC; otherwise, EC is dominant.

\begin{figure}
	\includegraphics[width=\columnwidth]{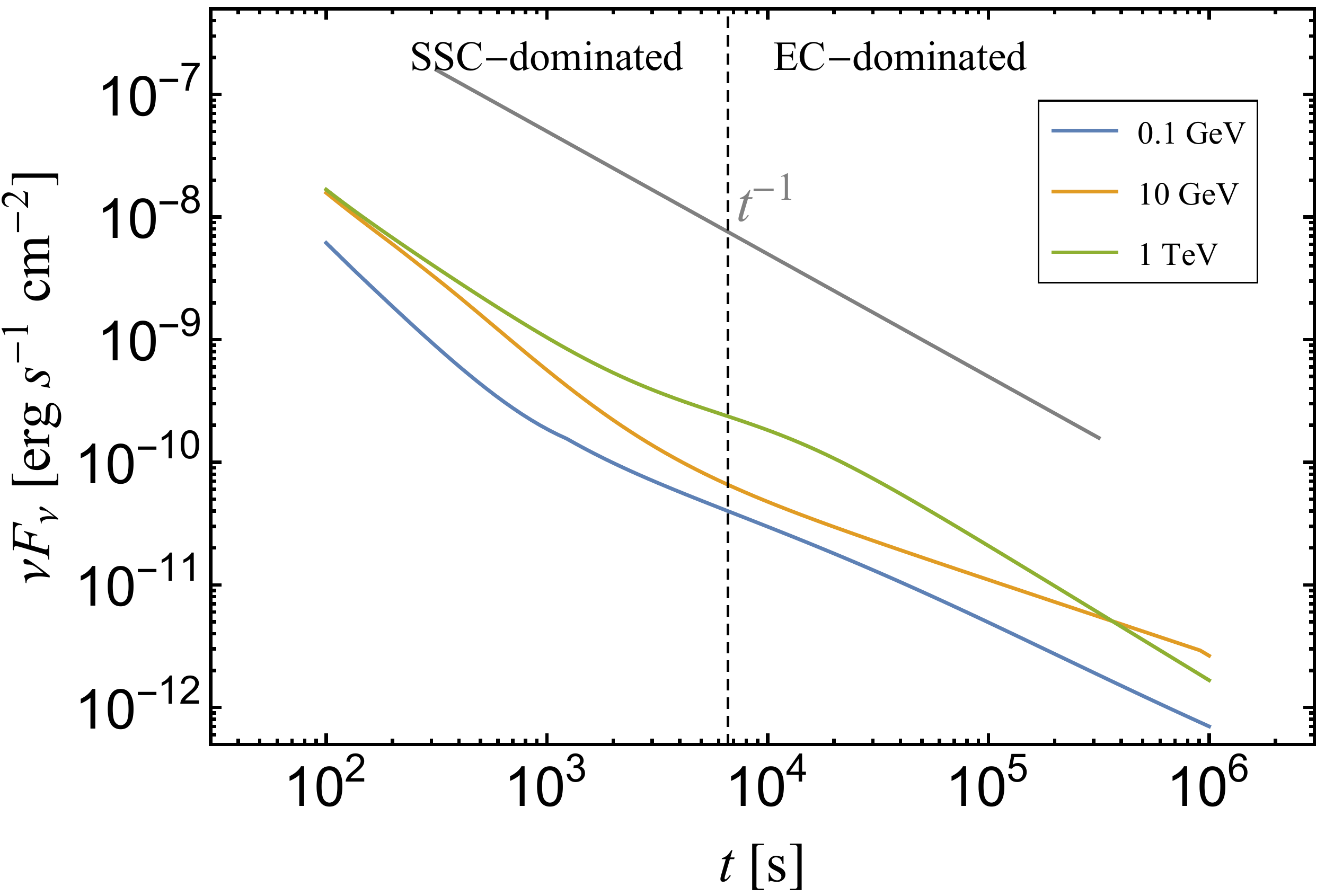}
	\caption{GRB afterglow light curves, from $0.1$~GeV to $1$~TeV, as produced from our analytical calculations for the parameters: $E=10^{54}$~erg, $A_{*}=0.3$, $\epsilon_{\rm e}=0.1$, $\epsilon_{\rm B} = 10^{-5}$, $p=2.2$, $U_{\rm ext}=7.5\times 10^{-6} {\rm erg\,cm^{-3}}$. The light curves transit from the SSC dominated regimes to EC dominated, divided by a dashed line at $t\approx 7000$~s. We also mark a temporal decay of $\sim t^{-1}$ resembling those found in GRB afterglow light curves by \fermi-LAT \citep{Ackermann2013}.
	A coloured version of this plot is available online. }
    \label{fig:lightcurves_all_wind}
\end{figure}

We also present parametric scalings of the observed inverse Compton flux on the model parameters for a wind-like medium. The flux of the EC component is given by
\begin{eqnarray}
\label{eqn:EC_flux_scaling_wind}
F_{\nu}\propto\left\{
 \begin{array}{ll}
E^{\frac{p+1}{2}} A^{-\frac{p-1}{2}} \epsilon_{\rm e}^{p-1} \nu^{-\frac{p-1}{2}} t^{-\frac{p-1}{2}} &  \nu^{\rm EC}_{\min}<\nu<\nu^{\rm EC}_{\rm c} \\
E^{\frac{p}{2}} A^{-\frac{p-2}{2}} \epsilon_{\rm e}^{p-1} \nu^{-\frac{p}{2}} t^{-\frac{p-1}{2}} & \nu>\nu^{\rm EC}_{\rm c}>\nu^{\rm EC}_{\min} \\
\end{array}
\right..
\end{eqnarray}
Similarly, the scaling for the SSC-dominated case reads
\begin{equation}
\label{eqn:SSC_flux_scaling_wind}
\resizebox{\linewidth}{!}{$F_{\nu}\propto\left\{
\begin{array}{ll}
E^{\frac{p-1}{2}}A^{\frac{-p+11}{4}}\epsilon_{\rm e}^{2(p-1)}\epsilon_{\rm B}^{\frac{p+1}{4}}\nu^{-\frac{p-1}{2}}t^{-p} & \nu^{\rm SSC}_{\min}<\nu<\nu^{\rm SSC}_{\rm c} \\ \\
E^{\frac{p}{2}}A^{\frac{p^2 - 14p + 24}{16 - 4p}}\epsilon_{\rm e}^{\frac{-2p^2+10p-8}{4-p}}\epsilon_{\rm B}^{\frac{-p^2+2p}{16-4p}}\nu^{-\frac{p}{2}}t^{\frac{p^2 - 3p}{4 - p}} & \nu>\nu^{\rm SSC}_{\rm c}>\nu^{\rm SSC}_{\min} \\
\end{array}
\right. .$}
\end{equation}

In Fig.~\ref{fig:lightcurves_all_wind} we plot, as an indicative example, the gamma-ray light curves (at three characteristic gamma-ray energies) produced via inverse Compton scattering from a blast wave propagating in a wind-like density environment. SSC dominates the electron cooling when $t<7000$~s, while EC becomes dominant at later times. The transition is marked by a dashed vertical line, and is related to a change in the temporal decay; the flux decays more slowly with time in the EC-dominated regime.

% Don't change these lines
\bsp	% typesetting comment
\label{lastpage}
\end{document}